\documentclass[superscriptaddress, aps, nofootinbib]{revtex4-2}

\usepackage{tabularx}
\usepackage{xcolor}
\usepackage{bm}
\usepackage{comment}
\usepackage{graphicx}
\usepackage{multirow}
\usepackage{array}
\usepackage{booktabs} 
\usepackage{tabularray}
\usepackage[normalem]{ulem}
\usepackage{amsmath}
\usepackage{mathtools}
\usepackage{amssymb}
\usepackage{acro}
\usepackage[utf8]{inputenc}
\usepackage[T1]{fontenc}
\usepackage{url}
\usepackage{verbatim}
\usepackage{mathrsfs}
\usepackage{mdframed}

\usepackage{newfloat}

\DeclareFloatingEnvironment[
    fileext=los,
    listname={Code Listings},
    name=Listing,
    placement=tbhp,
    within=none,
]{listing}

\DeclareUnicodeCharacter{0393}{$\Gamma$}

\usepackage[frozencache=true,cachedir=minted-cache]{minted}

\DeclareAcronym{DFT}{
short = DFT,
long = Density-Functional Theory
}

\DeclareAcronym{SOAP}{
short = SOAP,
long = Smooth Overlap of Atomic Positions
}

\DeclareAcronym{ACE}{
short = ACE,
long = Atomic Cluster Expansion
}

\DeclareAcronym{MACE}{
short = MACE,
long = Multi-ACE
}

\DeclareAcronym{RR}{
short = RR,
long =  Ridge Regression
}

\DeclareAcronym{KRR}{
short = KRR,
long = Kernel Ridge Regression
}

\DeclareAcronym{GAP}{
short = GAP,
long = Gaussian Approximation Potential
}

\newcommand{\revise}[1]{#1}
\newcommand{\revisetwo}[1]{#1}

\def\rr{{\bm r}}
\def\RR{\bm{R}}
\def\xx{{\bm x}}

\def\bAA{{\bm A}}

\def\ll{{\bm l}}
\def\nn{{\bm n}}
\def\mm{{\bm m}}
\def\kk{{\bm k}}

\def\zz{{\bm z}}

\def\BB{{\bm B}}

\def\cc{{\bm c}}

\def\given{\,|\,}

\def\calN{\mathcal{N}}
\def\numax{{\bar\nu}}

\begin{document}

\title{\revise{ACEpotentials}.jl : 
A Julia Implementation of the Atomic Cluster Expansion}

\author{William C Witt}
\affiliation{Department of Materials Science \& Metallurgy, University of Cambridge, Cambridge, United Kingdom}

\author{Cas van der Oord}
\affiliation{Engineering Laboratory, University of Cambridge, Cambridge, CB2 1PZ UK}

\author{Elena Gelžinytė}
\affiliation{Engineering Laboratory, University of Cambridge, Cambridge, CB2 1PZ UK}

\author{Teemu Järvinen}
\affiliation{Department of Mathematics, University of British Columbia, 1984 Mathematics Road, Vancouver, BC, Canada V6T 1Z2}

\author{Andres Ross} 
\affiliation{Department of Mathematics, University of British Columbia, 1984 Mathematics Road, Vancouver, BC, Canada V6T 1Z2}

\author{\revise{James P. Darby}}
\affiliation{Warwick Centre for Predictive Modelling, School of Engineering, University of Warwick, Coventry, CV4 7AL, United Kingdom}

\author{Cheuk Hin Ho} 
\affiliation{Department of Mathematics, University of British Columbia, 1984 Mathematics Road, Vancouver, BC, Canada V6T 1Z2}

\author{William J. Baldwin}
\affiliation{Engineering Laboratory, University of Cambridge, Cambridge, CB2 1PZ UK}

\author{Matthias Sachs}
\affiliation{School of Mathematics, University of Birmingham, Birmingham, B15 2TT, United Kingdom}

\author{James Kermode}
\affiliation{Warwick Centre for Predictive Modelling, School of Engineering, University of Warwick, Coventry, CV4 7AL, United Kingdom}

\author{Noam Bernstein}
\affiliation{Center for Materials Physics and Technology, U.~S. Naval Research Laboratory, Washington, DC, 20375, United States of America}

\author{G\'abor Cs\'anyi}
\email{Corresponding author, gc121@cam.ac.uk}
\affiliation{Engineering Laboratory, University of Cambridge, Cambridge, CB2 1PZ UK}

\author{Christoph Ortner}
\email{Corresponding author, ortner@math.ubc.ca}
\affiliation{Department of Mathematics, University of British Columbia, 1984 Mathematics Road, Vancouver, BC, Canada V6T 1Z2}

\date{\today}

\begin{abstract}
    We introduce {\tt \revise{ACEpotentials}.jl}, a Julia-language software package that \revise{constructs} interatomic potentials \revise{from quantum mechanical reference data} using the Atomic Cluster Expansion {\it (Drautz, 2019)}. \revise{As the latter provides a complete description of atomic environments, including invariance to overall translation and rotation as well as permutation of like atoms, the resulting potentials are systematically improvable and data efficient. Furthermore, the descriptor's expressiveness enables use of a linear model, facilitating rapid evaluation and straightforward application of Bayesian techniques for active learning.} We summarize \revise{the} capabilities \revise{of {\tt ACEpotentials.jl}} and demonstrate its strengths (simplicity, interpretability, robustness, performance) on a selection of prototypical atomistic modelling workflows.
\end{abstract}

\maketitle

\section{Introduction}
Machine-learning interatomic potentials (MLIPs) continue to revolutionize the fields of molecular and materials simulation~\cite{Behler2007ACSF, bartok2010gaussian}. MLIPs provide the means to simulate atomistic systems at or close to the accuracy of electronic structure methods, while being computationally cheaper by orders of magnitude. They make the simulation of large-scale systems and long time-scales at high model accuracy accessible and have therefore become an indispensable tool for atomic-scale simulation. Recent reviews of the the field are provided in \cite{2019_deringer_review, behler_csanyi_2021,DeringerCsanyi2021ChemRev,Ceriotti2021Review}. Of particular relevance to the present work are the methods introduced in \cite{bartok2010gaussian, THOMPSON2015SNAP, MTP2016, ACE_ralf}.

To create an MLIP, one begins with a flexible functional form, constrained only to comply with the natural symmetries of the potential energy in three-dimensional space, then estimates its parameters using reference data, typically in the form of energies, forces, and virial stresses for a set of representative atomic configurations. Ordinarily, the data are generated with quantum mechanical techniques, such as density functional theory calculations, which may be performed only for relatively small structures. 
A well-trained MLIP is then expected to provide accurate predictions of processes on similar but also much larger spatial scales. 

The {\it Atomic Cluster Expansion} (ACE) introduced in \cite{ACE_ralf} is a particular MLIP flavor that is flexible, theoretically well founded, interpretable, and for which it is straightforward to tune the cost-accuracy balance. It is \revise{establishing itself as} a successful MLIP approach for a wide range of tasks, especially but not exclusively in materials simulation; see e.g. \cite{2019_Seko, DUSSON2022, performant2022lysogorskiy, kovacs2021, hyperactive2022, 2023-ace_PtRh, 2022_pacemaker, 2022_drautz_ace_C}. Linear variants of the ACE model have been found remarkably data efficient and computationally efficient and as such have proven particularly useful for active learning (AL) workflows~\cite{hyperactive2022} as Sec.~\ref{sec:workflow} and Sec.~\ref{sec:computational_performance} will demonstrate. \revise{Linearity in particular enables sensitivity analysis and a path towards reliable uncertainty quantification.}

This article describes {\tt \revise{ACEpotentials}.jl}, which ties together a collection of Julia-language packages to expose a user-oriented interface facilitating the convenient construction of ACE MLIPs. 
To highlight the ease of use of our package, Listing~\ref{code:intro} provides a complete Julia-language example that produces an ACE potential for a TiAl dataset.

\begin{listing}[H]
\begin{mdframed}
\begin{minted}[xleftmargin=30pt,linenos]{julia}
using ACEpotentials
data, _, _ = ACEpotentials.example_dataset("TiAl_tutorial")
model = acemodel(elements = [:Ti, :Al],
                 order = 3,
                 totaldegree = 12,
                 Eref = [:Ti => -1586.0195, :Al => -105.5954])
acefit!(model, data)
export2lammps("TiAl_tutorial.yace", model)
\end{minted}
\end{mdframed}
\caption{A minimal Julia-language script for fitting an {\tt \revise{ACEpotentials}.jl} potential. It first downloads a training dataset, then uses {\tt acemodel} to create a model object whose parameters are explained fully in the following sections. The model parameters are estimated with the {\tt acefit!} command, and the result is exported in a LAMMPS compatible format.}
\label{code:intro}
\end{listing}

At the time of writing, {\tt \revise{ACEpotentials}.jl} provides interfaces for {\em linear} ACE models, which give good accuracy as well as performance both in parameter estimation and prediction. We have incorporated a range of geometric and analytical priors into the default model parameters that have proven robust in a range of tasks, including the challenging low data regime arising in active learning workflows. 
{\tt \revise{ACEpotentials}.jl} models can be used for molecular dynamics simulation in LAMMPS \cite{LAMMPS}, ASE  \cite{ase-paper} and {\tt Molly.jl} \cite{molly}. 

The Julia-language codes on which {\tt \revise{ACEpotentials}.jl} builds are written with ease-of-use, performance, and flexibility of model development in mind. Several variations and extensions of the ACE model implementations discussed in this article are under active development. \revise{The choice of Julia as the development language enables seamless transition from rapid prototyping to performance optimization. Moreover, Julia is establishing itself as leader in {\it scientific machine learning} (see, e.g., \cite{rackauckas2020universal}), facilitating highly customized model architectures with novel computational kernels.} 

Finally, we emphasize that the aim of this article is to illustrate the capabilities of {\tt \revise{ACEpotentials}.jl} but not to precisely document its use; for the latter see the reference material at \cite{ACEpotentials}, which will evolve along with the software. While the examples and code snippets provided throughout this article are compatible with the present version of \revise{{\tt ACEpotentials}.jl}, they should be taken primarily as illustrations of how the package may be used. The documentation will be kept up-to-date for the foreseeable future and will continually expand to describe additional options and features.

\section{Methods}
\label{sec:methods} 

\subsection{Review of the linear ACE framework}
\label{sec:model:ace_framework}
\subsubsection*{Model Specification}
An atomic structure is described by a collection of position-element pairs $(\rr_i, Z_i)$, and the computational unit cell (with open or periodic boundary conditions). In the ACE model, the total potential energy of such a structure is decomposed into site energies, 
\begin{equation}
    E = \sum_i \varepsilon_i, 
\label{eq:total_pot_energy}
\end{equation}
where the summation ranges over all atoms belonging to the computational cell and each $\varepsilon_i$ depends on its atomic neighbourhood containing all atoms within a cutoff radius $r_{\rm cut}$ from $\rr_i$, taking into account the boundary conditions. The ACE framework provides a design space to construct systematic models for the site energy $\varepsilon_i$ in terms of a complete linear basis of body-ordered symmetric polynomials.

For convenience we introduce the new variables $\xx_i := (\rr_i, Z_i)$ for the state of an atom and  $\xx_{ij} := (\rr_{ij}, Z_i, Z_j)$, where \revise{$\rr_{ij} = \rr_j - \rr_i$,} for the state of a bond between atoms $\xx_i, \xx_j$. In terms of these variables the site energy is expanded in body-order, in two different formulations:  
\begin{subequations} \label{eq:manybody}
    \begin{align} 
     \label{eq:manybody_pure}
    \varepsilon_i
    &= 
    V^{(0)}(Z_i) + \sum_{j_1} V^{(1)}(\xx_{i j_1})
    + \sum_{j_1 < j_2} V^{(2)}(\xx_{i j_1}, \xx_{i j_2}) 
    + \cdots + 
    \sum_{j_1 < \cdots < j_{\numax}} 
    V^{(\numax)}(\xx_{i j_1}, \dots, \xx_{i j_{\numax}}) \\
     \label{eq:manybody_dirty}    
    &= 
    V^{(0)}(Z_i) + \sum_{j_1} U^{(1)}(\xx_{i j_1})
    + \frac{1}{2!} \sum_{j_1, j_2} U^{(2)}(\xx_{i j_1}, \xx_{i j_2}) 
    + \cdots + 
    \frac{1}{\numax!} \sum_{j_1,\dots, j_{\numax}} 
    U^{(\numax)}(\xx_{i j_1}, \dots, \xx_{i j_{\numax}}).
    \end{align} 
\end{subequations}
We call the first formulation \eqref{eq:manybody_pure}  the {\em canonical cluster expansion}. It can be transformed \cite{ACE_ralf} into the second formulation \eqref{eq:manybody_dirty}, where the sums run over all possible combinations of atoms, including all permutation-equivalent clusters and even ``artificial clusters'' with repeated particles. This \revise{transformation} introduces unphysical self-interaction terms such as $V^{(2)}(\xx_{ij}, \xx_{ij})$, but this counter-intuitive choice leads to a tensor product structure that can be exploited in constructing a highly efficient evaluation scheme. Our code is unique in that it implements the transformation between the two descriptions and also allows the evaluation of the canonical formulation \eqref{eq:manybody_pure}. Indeed the default {\tt \revise{ACEpotentials}.jl} model specification uses a combination of the two formulations.
We will briefly review the challenges involved in evaluating cluster expansion models in \revise{Appendix~\ref{sec:implementation}.}

Both series in \eqref{eq:manybody} are truncated versions of an exact body-order expansion. An exact expansion would include terms up to the number of atoms in the system, while here the maximum body-order is $\numax+1$ (corresponding to a correlation order of $\numax$), which constitutes the first approximation parameter. In practice, the truncation is performed at low to moderate $\numax$ (typically $5$ or less) for several reasons, including control of model complexity and computational cost. 

Each potential $V^{(\nu)}$ (or, $U^{(\nu)})$ is parameterized by a linear model, a process for which we give details below in the following sections. This then results in a parameterisation of the site energy that is also linear, 
\begin{equation}
    \label{eq:lin_ace}
    \varepsilon_i = \cc \cdot \BB_i,
\end{equation}
where $\cc$ is a vector of parameters and $\BB_i$ a vector of basis functions (or, features) involved in the expansion of the many-body potentials $V^{(\nu)}$ or $U^{(\nu)}$. The basis functions are by construction invariant under rotations, reflections and permutations of like atoms. The representation is also {\em complete} (or, universal) in the sense that when the approximation parameters (body-order, cutoff radius, and expansion resolution) are taken to infinity, the model can in principle represent an arbitrary smooth site-energy potential. Linearity of the model allows us to employ a vast range of established tools for parameter estimation and uncertainty quantification, and enables rapid model development by refitting to new training data or with adjusted hyperparameters.

The basis functions $\BB_i$ specify the model. In a \revise{typical} example this can be done as demonstrated in Listing~\ref{code:acemodel}. 

\begin{listing}[H]
\begin{mdframed}
\begin{minted}[xleftmargin=30pt, linenos=true]{julia}
using ACEpotentials
model = acemodel(; elements = [:Ti, :Al], 
                   order = 3,
                   totaldegree = 12, 
                   rcut = 5.5,
                   Eref = [:Ti => -1586.0195, :Al => -105.5954])
\end{minted}

\qquad \begin{minipage}{0.8\textwidth}
\begin{tabular}{rc|cl}
    {\tt elements} &&& list of chemical elements occurring in the system of interest \\ 
    {\tt order} &&& maximum correlation order, $\numax$ in the article text; cf. Eq. \eqref{eq:manybody} \\ 
    {\tt totaldegree} &&& spatial resolution of the $\nu$-body potentials; cf. Eq. \eqref{eq:total_degree} \\ 
    {\tt rcut} &&& (optional) cutoff radius; cf. Sec.~\ref{sec:model:basis} \\
    {\tt Eref} &&& (optional) reference energies specifying $V^{(0)}(Z_i)$
\end{tabular}
\end{minipage}
\end{mdframed}
\caption{A typical construction of an ACE model and description of parameters. }
\label{code:acemodel}
\end{listing}

The {\tt model} object specifies the model site energy potential, from which derived properties such as potential energy, forces and virial stresses can be computed that are used in molecular statics, molecular dynamics or sampling algorithms.

There are many additional parameters and options available to specify an ACE model, some of which we discuss throughout the remainder of this paper. For a complete list of options we refer to the documentation \cite{ACEpotentials}. We only remark briefly on the {\tt Eref} parameter:  We recommend the explicit specification of the one-body term $V^{(0)}$. We observed in many tests that constraining $V^{(0)}(Z_i)$ to be the energy of a single isolated atom with atomic number $Z_i$ yields more chemically realistic potentials that are more robust in practical molecular dynamics and molecular statics simulations, especially those involving breaking and forming bonds. One provides this information to an ACE model as shown in Listing~\ref{code:acemodel}, line 6.

\revise{In the remainder of this section we maintain a focus on high level intuitive understanding of options and parameters and avoid details and technicalities of the ACE framework as much as possible. For those details we refer to Appendix~\ref{sec:implementation} and to the many publications now available on the subject \cite{ACE_ralf, DUSSON2022, Ceriotti2021Review, performant2022lysogorskiy}.}

\subsubsection*{Parameter Estimation}
Having specified a physically reasonable model architecture, we must now estimate its parameters. To that end we require a training set, which typically consists of a list of atomic structures, $\RR = \{ R \}$, for which the total potential energy $\mathscr{E}_R \in \mathbb{R}$, forces $\mathscr{F}_R \in \mathbb{R}^{3 \times N_R}$ (with $N_R$ the number of atoms in the computational unit cell) and possibly also virial stresses \revise{$\mathscr{V}_R \in \mathbb{R}^{6}$ (in Voigt notation)} have been evaluated with an electronic structure model. We define $E(\cc; R), F(\cc; R), V(\cc; R)$ be the corresponding energies, forces and virials for the structure $R$ in the ACE model, with parameters $\cc$. The simplest way to estimate those parameters is then to minimize the least squares loss function 
\begin{equation} \label{eq:lsq-loss}
\begin{split}
    L(\cc) &= \sum_{R \in \boldsymbol{R}} \Big( w^{2}_{E,R} | E(\cc; R) - \mathscr{E}_{R} |^{2} + w^{2}_{F,R} | F(\cc; R) - \mathscr{F}_{R} |^{2} + w^{2}_{V,R}| V(\cc; R) - \mathscr{V}_{R} |^{2} \Big).
\end{split}
\end{equation}
The weights $w_{E,R}, w_{F,R}, w_{V,R}$ can be used to give more or less relative ``importance'' to certain structures or observations. They are usually highly structured (e.g., $w_{E,R}, w_{V,R}$ are scaled with the number of atoms in a structure $R$), which will be discussed in more detail in \revise{Section~\ref{sec:parameter_estimation}.}
Since the ACE model is linear in $\cc$ it follows that $L(\cc)$ is quadratic, which means that minimizing $L$ is a linear least squares problem. A wide range of efficient numerical techniques exist for its solution. In particular we will normally employ regularized or Bayesian variations of the naive least squares minimization, which are discussed in \revise{Sections~\ref{sec:parameter_estimation} and~\ref{sec:bayes}.}

In Listing~\ref{code:acefitsimple} we read in such a prepared training set provided in the extended XYZ format and then estimate the model parameters with a default solver (Bayesian Linear Regression; cf. \revise{Section~\ref{sec:bayes}}). Several steps are combined and hidden from the user in the {\tt acefit!} convenience function, but all these steps can in principle also be performed manually, e.g., to explore different parameter estimation algorithms that are currently not interfaced by {\tt \revise{ACEpotentials}.jl}. In line 5 of the listing, the fitted model is exported to a format that can be used for molecular dynamics simulations in LAMMPS.

\begin{listing}[H]
\begin{mdframed}
\begin{minted}[escapeinside=||, xleftmargin=30pt, linenos=true]{julia}
model = ... # cf. Listing |\ref{code:acemodel}|
P = smoothness_prior(model)
data, _, _ = ACEpotentials.example_dataset("TiAl_tutorial")
acefit!(model, data; prior = P, solver = ACEfit.BLR())
export2lammps("TiAl.yace", model)
\end{minted}

\qquad \begin{minipage}{0.8\textwidth}
\begin{tabular}{rc|cl}
    {\tt smoothness\_prior} &&& specifies a model prior / regularizer; cf. \revise{Section~\ref{sec:basis_selection}} \\
    {\tt pathtodata} &&& absolute path to a small training set used for testing \\ 
    {\tt data} &&& collection of structures containing training data; cf. \revise{Section~\ref{sec:data}} \\ 
    {\tt acefit!} &&& assembles and solves the least squares system; cf. \revise{Section~\ref{sec:parameter_estimation}} \\ 
    {\tt ACEfit.BLR()} &&& default solver for parameter estimation; cf. \revise{Section~\ref{sec:bayes}} \\ 
    {\tt export2lammps} &&& exports the model to a LAMMPs readable format.
\end{tabular}
\end{minipage}
\end{mdframed}

\caption{A representative example loading a training dataset and estimating ACE model parameters.}
\label{code:acefitsimple}
\end{listing}

In the remainder of \revise{Section~\ref{sec:methods}} we will dive slightly deeper into some the steps we outlined above. Then, in \revise{Section~\ref{sec:workflow}} we will demonstrate how the framework can be used to fit potential energy models for realistic materials and molecular systems of scientific interest. 

\subsection{Choice of basis functions \& Geometric priors}
\label{sec:model:basis}
The parameters in the model specification in Listing~\ref{code:acemodel} specify a basis in which the $V^{(\nu)}$ potentials are expanded. In the current section we will detail the {\em basis functions} that are employed, while in \revise{Section~\ref{sec:basis_selection}} we will then explain how to select a finite subset from the infinite complete basis set.

\subsubsection*{One-particle basis}
To begin we must select a {\it one-particle basis} $\phi_k$ in which all smooth functions $f(\xx_{ij}) = f(\rr_{ij}, Z_i, Z_j)$ can be expanded. The most general form we consider is
\begin{equation}
    \phi_{znlm}(\rr_{ij}, Z_i, Z_j) 
    = 
    R_{nl}(r_{ij}, Z_i, Z_j) Y_l^m(\hat{\rr}_{ij}) \delta_{z Z_j},
\end{equation}
\revise{where $\delta$ denotes the Kronecker symbol and we have identified $k = (z,n,l,m)$.}
The $Y_l^m$ are the standard complex spherical harmonics, while $R_{nl}$ is called the {\em radial basis}. The choice of $Y_l^m$ to embed the angular component $\hat\rr_{ij}$ facilitates the exact symmetrization of the parameterisation with respect to rotations. Since $(r_{ij}, Z_i, Z_j)$ is already invariant under rotations, the choice of $R_{nl}$ is extremely general. Nevertheless we will below outline a heuristic that leads to a narrow class of choices that have proven successful in many applications. However, we note that the optimal choice of $R_{nl}$ remains an active area of research and will likely also evolve within {\tt \revise{ACEpotentials}.jl}.

Once $\phi_k$ is selected, each potential $V^{(\nu)}$ \revise{(or, $U^{(\nu)}$)} is expanded in terms of a tensor product many-body basis, 
\begin{equation} \label{eq:expansion_Vnyu}
    \begin{split}
        V^{(1)}\big( \xx_{i j_1} \big) &= 
         \sum_{k_1} c^{(Z_i)}_{k_1} \phi_{k_1}(\xx_{i j_1}) 
        \\
        V^{(2)}\big( \xx_{i j_1}, \xx_{i j_2} \big)
        &=
        \sum_{k_1, k_2} c_{k_1 k_2}^{(Z_i)} \phi_{k_1}\big( \xx_{ij_1} \big) \phi_{k_2}\big( \xx_{i j_2} \big) \\
        \vdots \qquad & \qquad \qquad \vdots \\
        V^{(\numax)}\big( \xx_{i j_1}, \dots, \xx_{i j_{\numax}} \big)
        &=
        \sum_{k_1, \dots, k_{\numax}}c_{k_1\cdots k_{\numax}}^{(Z_i)} \phi_{k_1}\big( \xx_{ij_1} \big) \cdots \phi_{k_{\numax}}\big( \xx_{i j_{\numax}} \big)
    \end{split}
\end{equation}
The model parameters $c^{(Z_i)}_{k_1 \cdots k_\nu}$ will be estimated from data. Note that we choose individual model parameters for each center-atom element $Z_i$. During the parameter estimation, the parameters will be constrained to guarantee invariance of the resulting potentials under rotations and reflections of an atomic environment. Invariance under permutations is already ensured through the summation in \eqref{eq:manybody}. \revise{Appendix~\ref{sec:implementation} reviews additional details of this invariant basis construction, resulting in the specification of $\BB_i$ in terms of which site energy is defined in \eqref{eq:lin_ace}.}

To complete the model specification two steps remain: \revise{(i)} the choice of radial basis $R_{nl}$; and \revise{(ii)} the selection of basis functions $(k_1, \dots, k_\nu)$ that we employ in the expansions \eqref{eq:expansion_Vnyu}. In the remainder of this section we discuss \revise{(i) while (ii)} will be discussed in \revise{Section~\ref{sec:basis_selection}.} 

\subsubsection*{Radial basis} 
There is considerable freedom in the choice of the radial basis $R_{nl}$, which can be thought of as a {\em geometric prior}. For example, it incorporates the interaction range (cutoff radius, $r_{\rm cut}$) and can be tuned to capture rough qualitative information about interacting atoms. In the following we describe a class of radial bases, available through {\tt \revise{ACEpotentials}.jl}, that require no data-driven optimization and thus leads to genuinely linear models. \revise{At the time of writing this article, {\tt \revise{ACEpotentials}.jl} supports radial bases indexed by $n$ only, i.e. $R_{nl} = R_n$ for all $l$.} This class is described by 
\begin{equation}
    R_{n}(r_{ij}, Z_j, Z_i)
    = f_{\rm env}(r_{ij}, Z_j, Z_i) P_n\big( y(r_{ij}, Z_j, Z_i) \big),
\end{equation}
with the following components: 
\begin{itemize}
    \item $y$ is an element-dependent distance transform, which can be used to impose increased spatial resolution where needed, especially near the equilibrium bond-length. \revise{We typically employ
    \[
        y(r_{ij}, Z_i, Z_j) = 
        \bigg(1 + a \frac{(r/r_0)^q}{1 + (r/r_0)^{q-p}}\bigg)^{-1}, 
    \]
    where $r_0$ is an estimate of the equilibrium bond-length in the system and $a$ is chosen to maximize the gradient of $y$ at $r = r_0$, thereby maximizing resolution for nearest-neighbour interaction. 
    The idea behind this transform  is that it behaves as $r^{-q}$ for large $r$ and as $1-r^p/a$ for small $r$ thereby decreasing resolution in those two limits at rates determined by the parameters $p, q$. The reduction in resolution in the small $r$ regime is desirable when no data is available to specify the model in that regime; see also Figure~\ref{fig:transforms}. }

    \begin{figure}
        \includegraphics[height=7cm]{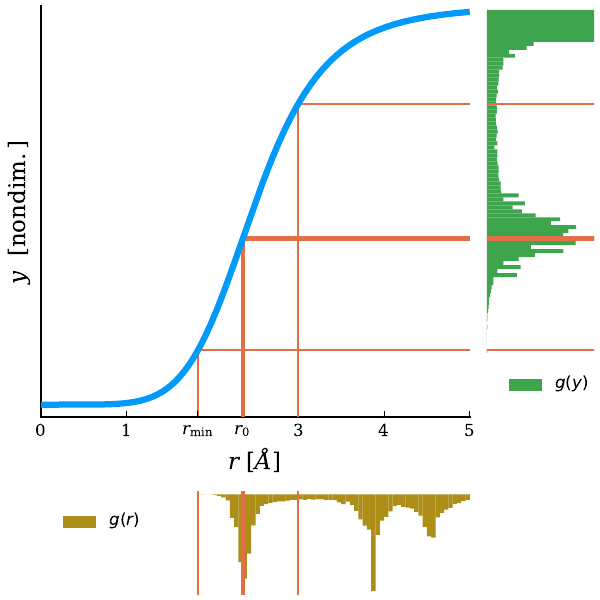}
        \includegraphics[height=7cm]{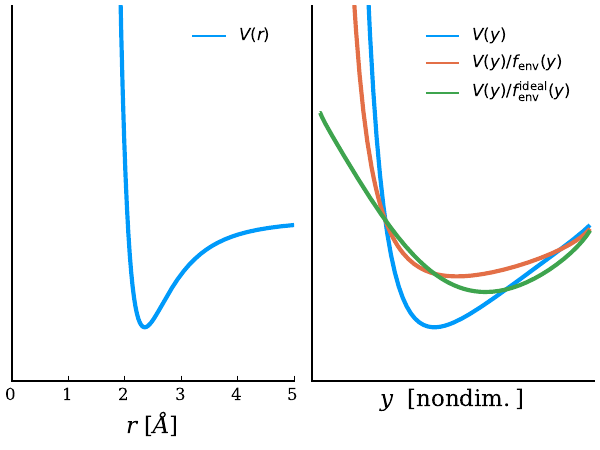}
        
        \caption{ \label{fig:transforms} {\bf Center:} a typical interaction potential $V(r)$, plotted in $r$-coordinates. {\bf Left:} a coordinate transform $y = y(r)$ \revise{to a non-dimensional variable $y$} that increases resolution near $r = r_0$ where the potential minimum is located and decreases resolution \revise{below $r_{\rm min}$ (the radial distance occuring in the training dataset)}, to zero near $r = 0$ where there is no data (and the envelope $f_{\rm env}$ becomes relevant) and near $r = r_{\rm cut}$ where the potential converges to a constant. The histograms show the distribution of a typical dataset in \revise{both} $r$\revise{- and } $y$-coordinates.  
        {\bf Right:} the interaction potential plotted (i) in transformed coordinates $V(r(y))$, (ii) with the default \revise{pair} envelope removed and (iii) with the theoretically optimal, typically unknown, envelope removed. 
        The parameterisation and the smoothness priors are not applied to the original potential $V(r)$ but to the transformed potential $V(y) / \revise{f_{\rm env}(y)}$.}
    \end{figure}

    \item $P_n$ is an orthogonal basis in $y$-coordinates. Our default choice is the Legendre orthogonal polynomial basis, which implicitly assumes equidistribution of resolution in $y$-coordinates. 
    \item Finally, $f_{\rm env}$ is an envelope that specifies the cutoff radius $r_{\rm cut}$.
    \begin{itemize}
    \item The default and canonical choice for the many-body basis is 
    \[
            f_{\rm env}(r_{ij}, Z_i, Z_j) = y^2 (y - y_{\rm cut})^2,
    \]
    \revise{where 
    $y_{\rm cut} = y(r_{\rm cut}, Z_i, Z_j)$.}

    \item The default choice of envelope for the pair potential $U^{(1)}$ or $V^{(1)}$ is Coulomb potential tilted to ensure a smooth cutoff, 
    \[
            f_{\rm env}(r_{ij}, Z_i, Z_j) = 
            {\textstyle \big(\frac{\revise{r_{ij}}}{r_0})^{-1} 
                    - \big(\frac{r_{\rm cut}}{r_0}\big)^{-1} 
                    + \big(\frac{r_{\rm cut}}{r_0}\big)^{-2}
                    \big( \frac{\revise{r_{ij}}}{r_0} - \frac{r_{\rm cut}}{r_0} \big), }
    \]
    which is repulsive as \revise{$r_{ij}^{-1}$} as $r \to 0$ but continuously differentiable at the cutoff.
    \end{itemize}
    
    While the envelope for the many-body potential is canonical, for the pair potential envelope there is significant scope for inserting prior modelling knowledge of the system of interest. For example, one could replace the $r^{-1}$ type behaviour with $r^{-p} + r^{-q}$ to obtain different behaviour as $r \to 0$ and $r \to r_{\rm cut}$, or in fact one could incorporate the ZBL potential \revise{\cite{ziegler_stopping_1985}} to obtain asymptotically exact repulsion. 

    The effect of the distance transform $y = y(r)$ and of the envelope function are visualized in Figure~\ref{fig:transforms}.

    \item Repulsion restraint: The construction outlined above means that, in the canonical cluster expansion formulation, the pair potential is given by 
    \[
        V^{(1)}(r_{ij}, Z_i, Z_j) = f_{\rm env}(r_{ij}, Z_i, Z_j) p_{Z_i Z_j}(y_{ij}),
    \]
    where $p_{Z_iZ_j}$ is a polynomial in transformed $y$ coordinates. By imposing the constraint that $p_{Z_i Z_j}(y_0) = 1$, where $y_0 = y(0, Z_i, Z_j)$, we ensure that $E \sim f_{\rm env}(r_{ij})$ as $r_{ij} \to 0$. This {\em guarantees} repulsive behaviour of the total energy, independently of whether or not this is provided through the training data. In practice we enforce this weakly through a mild restraint to give the potential more flexibility.
\end{itemize}

\begin{listing}[H]
\begin{mdframed}
    \begin{minted}[escapeinside=||, xleftmargin=30pt, linenos=true]{julia}
    using ACEpotentials
    elements = [:Ti, :Al]
    totaldegree = 12
    r0 = (rnn(:Ti) + rnn(:Al)) / 2
    rcut = 2 * r0
    trans = AgnesiTransform(; r0=r0, p = 2)
    fenv = PolyEnvelope(1, r0, rcut)
    radbasis = transformed_jacobi_env(totaldegree, trans, fenv, rcut)
    model = acemodel(elements = elements,
                     order = 3,
                     totaldegree = totaldegree,
                     radbasis = radbasis)
    \end{minted}
\end{mdframed}
\caption{A example demonstrating more fine-grained control over the choice of radial basis $R_{nl}$. The function {\tt transformed\_jacobi\_env} constructs the polynomial basis from which the radial basis is constructed, which can be within the general class of Jacobi polynomials, but is normally taken to be the Legendre basis in transformed $y$ coordinates.}

\label{code:radialbasis}
\end{listing}

\subsection{A priori sparsification \& Smoothness prior}
\label{sec:basis_selection}
We now turn towards the second aspect of basis construction: how to select which of the infinitely many tensor product basis functions 
\begin{equation} \label{eq:tensorbasfcn}
    \phi_{k_1} \otimes \cdots \otimes \phi_{k_\nu}, 
\end{equation}
specified by the tuples $(k_1, \dots, k_\nu)$, we wish to incorporate into the expansion of the $(\nu+1)$-body potential $V^{(\nu)}$. 

\subsubsection*{Sparse basis selection}
Recall that $k_t = (z_t, n_t, l_t, m_t)$, and that the bound  $|m_t| \leq l_t$ on $m_t$ automatically gives a selection of possible $m_t$ values once $l_t$ bounds are chosen. Roughly speaking, $n_t, l_t$ measure how oscillatory the corresponding basis functions are in, respectively, the radial $r_t$ and angular $\hat\rr_t$ coordinates. Therefore one typically puts upper bounds $n_t \leq n_{\rm max}$ and $l_t \leq l_{\rm max}$ in the basis selection, i.e. one chooses all basis functions $(k_1, \dots, k_\nu)$ in the expansion for which these bounds are satisfied. Lower bounds lead to a smaller basis, but also less flexibility and correspondingly lower accuracy on the training set.

This simple strategy is available in {\tt \revise{ACEpotentials}.jl} but the default usage takes the notion of regularity a step further and bounds the {\em mixed regularity} of the basis functions we select. \revise{This is done by choosing a maximum {\em total} degree ${\rm totaldegree}(\nu)$ for each correlation order $\nu$} and choosing all basis functions $(k_1, \dots, k_\nu)$ such that 
\begin{equation} \label{eq:total_degree}
    1 \leq \nu \leq \numax \quad \text{and} \quad 
    \sum_{t = 1}^\nu n_t + w_{\rm L} l_t \leq {\rm totaldegree}(\nu). 
\end{equation}
The additional weight $w_{\rm L}$ allows us to select whether we require lower or higher resolution of the angular versus radial components of the interaction. Note that a higher weight $w_{\rm L}$ decreases the angular resolution. The resulting selected basis is much sparser and is appropriate for parameterising very smooth functions in high dimension.

The default usage is that \revise{${\rm totaldegree}(\nu)$} takes the same value for all $\nu$ but one may also specify a separate total degree for each correlation order $\nu$. For example, Listing \ref{code:acefitsparse} demonstrates how to select a stronger weight $w_{\rm L} = 2.0$ thus providing less angular resolution, as well as how to select total polynomial degrees $25, 23, 20, 10$ for, respectively, parameterising $V^{(1)}, V^{(2)}, V^{(3)}, V^{(4)}$. 

\begin{listing}[H]
\begin{mdframed}
\begin{minted}[escapeinside=||, xleftmargin=30pt, linenos=true]{julia}
using ACEpotentials
model = acemodel(elements = [:Ti, :Al],
                 order = 4,
                 wL = 2.0,
                 totaldegree = [25, 23, 20, 10])
\end{minted}

\qquad \begin{minipage}{0.8\textwidth}
\begin{tabular}{rc|cl}
    {\tt wL} &&& specifies the relative resolution in angular and radial basis \\ 
    {\tt totaldegree} &&& specify seperate degrees for each correlation order
\end{tabular}
\end{minipage}
\end{mdframed}
\caption{Construct an ACE model with finer control on the sparse selection of basis functions.}
\label{code:acefitsparse}
\end{listing}

Significant further fine-tuning of the basis specification is possible, e.g.\ choosing different total degrees and $w_{\rm L}$ parameters for different interacting species. This is explained in the package documentation \cite{ACEpotentials}.

\subsubsection*{Smoothness Prior}
The foregoing discussion concludes the model {\em architecture} specification. An issue closely related to the sparse basis selection \eqref{eq:total_degree} is the definition of a smoothness prior that may be employed for ridge regression (regularized least squares) which we discuss in \revise{Section~\ref{sec:parameter_estimation}} or in the Bayesian framework of \revise{Section~\ref{sec:bayes}}. As explained above, the value 
\[
    \sum_{t = 1}^\nu n_t + w_{\rm L} l_t
\]
is a qualitative estimate for how oscillatory or smooth a basis function \eqref{eq:tensorbasfcn} is. We can extend this definition slightly by adding another parameter $p$ and defining 
\begin{equation} \label{eq:smoothness_prior}
    \revise{\gamma_{\zz \nn \ll \mm}} := \sum_{t = 1}^\nu n_t^p + w_{\rm L} l_t^p,
\end{equation}
\revise{where $\zz = (z_t)_{t=1}^\nu, \nn = (n_t)_{t=1}^\nu, \ll = (l_t)_{t=1}^\nu$ and $\mm = (m_t)_{t=1}^\nu$.} 
We then collect these parameters into a diagonal matrix $\Gamma$ with $\Gamma_{\kk \kk} = \gamma_{\kk}$. If $\cc$ are the model parameters then $\| \Gamma \cc \|_2$ will be a rough estimate for how smooth the potential energy surface is.

The matrix $\Gamma$ also serves as a smoothness prior within the Bayesian interpretation of ridge regression: the prior distribution for the model parameters $\cc$ is given by \revise{a multivariate normal distribution that is centered at the origin and has variance proportional to $\Gamma^{-2}$;
see Sections~\ref{sec:parameter_estimation}~and~\ref{sec:bayes}}. In {\tt \revise{ACEpotentials}.jl} this operator can be constructed as shown in Listing~\ref{code:smoothnessprior}, with $p = \revise{4}, w_{\rm L} = 1$ the default.

\begin{listing}[H]
\begin{mdframed}
\begin{minted}[escapeinside=||, xleftmargin=30pt, linenos=true]{julia}
model = ... # cf. Listing |\ref{code:acemodel}|
Γ = smoothness_prior(model; p = 4, wL = 1)
\end{minted}

\end{mdframed}
\caption{Construct an operator that estimates the smoothness of the MLIP model, to be used as a \revise{Tikhonov} regulariser, or prior in a Bayesian framework.}
\label{code:smoothnessprior}
\end{listing}

The resulting operator $\Gamma$ may now be used to specify the regularizer (or prior) of parameter estimation algorithms, e.g., in Listing~\ref{code:acefitsimple}, line 2 and explained in more detail in \revise{Sections~\ref{sec:parameter_estimation} and~\ref{sec:bayes}.} A key point is that $\Gamma$ is a {\it rigorous} smoothness prior for the canonical cluster expansion \eqref{eq:manybody_pure} but only a heuristic for the self-interacting expansion \eqref{eq:manybody_dirty}.

\revise{It is interesting in general, but in particular in the low-data regime, to explore different choices of priors. Two particular variants that are also available in {\tt ACEpotentials.jl} are the exponential and Gaussian priors 
\[
    \gamma_{\zz \nn \ll \mm}^{\rm exp}
    = \exp\Big( \alpha_{\rm l} \sum_t l_t + \alpha_{\rm n} \sum_t n_t \Big), 
    \qquad \text{and} \qquad 
    \gamma_{\zz \nn \ll \mm}^{\rm gauss}
    = 
    \exp\Big( 
        \sigma_{\rm l} \sum_t l_t^2 
        + \sigma_{\rm n} \sum_t n_t^2 
    \Big), 
\]
which enforce even stronger smoothness requirements than the algebraic prior \eqref{eq:smoothness_prior} and are currently still experimental features. 
}

\subsection{Training data}
\label{sec:data}
In the foregoing sections we discussed in some depth how an ACE interatomic potential architecture can be conveniently specified. The next task is to estimate the parameters matching the model to training data. 

A training dataset consists of a collection of reference structures, $\RR = \{ R \}$, each with associated potential energy $\mathscr{E}_R \in \mathbb{R}$, forces $\mathscr{F}_R \in \mathbb{R}^{3 \times N_R}$ and, when appropriate, virials \revise{$\mathscr{V}_R \in \mathbb{R}^{6}$ (Voigt notation)}. The reference energies, forces and virials are typically obtained by evaluating a ``high fidelity'' reference potential energy surface for which we wish to obtain an ACE surrogate model. Density Functional Theory is a common choice, but higher levels of theory such as Coupled-Cluster methods are also used, especially for non-periodic systems. In addition each training structure should be given a label that specifies related sub-groups. For example, these subgroups could indicate different phases of a material, and the resulting labels might be ``bcc'', ``fcc'', ``liquid''. The label could also indicate the MD temperature from the which the structures were generated, e.g. ``fcc500K'' or ``liquid2500K''. This allows convenient filtering of the training set, e.g., for assigning training weights (cf. \revise{Section~\ref{sec:parameter_estimation}}) or fitting to subsets. 

Acquisition of training data need not be performed within the {\tt \revise{ACEpotentials}.jl} package, but can be undertaken in any simulation software that makes it convenient to generate and manipulate atomic structures, perform molecular dynamics or Monte Carlo simulations, and to evaluate structures using a high fidelity electronic structure model. Because of the general ease of use and in particular ease of interoperability with the Julia molecular simulation eco-system, we often use the Atomic Simulation Environment~\cite{aseLarsen2017}.  

The standard format for storing and retrieving a training set in {\tt \revise{ACEpotentials}.jl} is the extended XYZ format and can be read as shown in Listing~\ref{code:readextxyz}. This results in a list of atomic structures storing the structure information as well as the training data.

\begin{listing}[H]
\begin{mdframed}
\begin{minted}[breaklines, escapeinside=||, xleftmargin=30pt, linenos=true]{julia}
using ACEpotentials
pathtodata = "path/to/data.xyz"
data = read_extxyz(pathtodata)
\end{minted}

\end{mdframed}
\caption{Reading a training set from an extended XYZ file. }
\label{code:readextxyz}
\end{listing}

\subsubsection*{Overview of Training Set Acquisition}

The acquisition of training data is often the most time-consuming aspect of MLIP development. An in-depth discussion goes beyond the scope of this software review article; important details can be found for example in \cite{Szlachta2014-qr,Bartok2018-fk,DeringerCsanyi2021ChemRev,performant2022lysogorskiy,Bartok2017-al}.
In the remainder of this section we give an outline of general strategies to consider, while in Section~\ref{sec:workflow} we go into practical aspects how training sets can be constructed in a few prototypical applications and what kind of tools {\tt \revise{ACEpotentials}.jl} provide to support that task. 

The overarching requirements are that training sets (1) must contain small enough atomic structures that they can be evaluated using high-fidelity electronic structure models; and (2) must contain snapshots of all possible local atomic configurations one expects to encounter during simulation and prediction tasks. Thus, generating a training set reduces to generating representative atomic structures which are then evaluated with the reference model to obtain target potential energies, forces and virials. While the latter is usually straightforward and varies little between projects, there is no standard way yet to generate the training structures. The choice will depend on the atomic system at hand, and the simulation tasks that the model must be able to perform reliably, e.g. which system properties (observables) are to be modelled. 

As a first step, one should ``sketch out'' the parts of the potential energy landscape that are of interest, e.g. construct one representative structure for each distinct energy minimum of interest. This might include different phases or material defects that the final model should be able to describe. Next, one generates random samples from those sketches for example by displacing the atom positions (randomly, along normal modes, volume scans, and so forth), or by subsampling an {\it ab initio} molecular dynamics trajectory. After collecting a seemingly adequate number of training structures (the total number of observations should normally exceed that number of parameters) one can fit a first model and test that model's accuracy with respect to some target property. If the accuracy is inadequate, or the model not robust (e.g., an MD simulation is unstable), then a good strategy is to proceed with an iterative model refinement process. In each iteration additional training structures are selected to converge the model's accuracy with respect to the target properties of interest. One might add hand-crafted structures to fix a particular flaw (e.g. to improve description of inter-molecular interaction in a molecular liquid or include supercells with vacant atomic sites) or model-driven MD to less computationally expensively explore relevant parts of Potential Energy Surface (for example, low potential energy regions to bring potential-Boltzmann-sample closer to reference-Boltzmann-sample and wider temperature/pressure range than intended for application of interest to make the model-driven simulations more stable). 

Iterative model refinement is closely related to \emph{active learning}. That strategy assumes that there is an accurate and efficient way available to estimate model uncertainty. During a simulation task, for example a molecular dynamics simulation, when a structure with high uncertainty is encountered it is evaluated with a reference method and added to the training data. To accelerate this process, we developed Hyper-Active Learning~\cite{hyperactive2022}, which biases molecular dynamics simulation towards high-uncertainty and high predicted error regions. This strategy is sometimes capable of more rapidly generating many independent training samples. Section~\ref{sec:workflow} will go into some details how this strategy is used in practice.

\subsection{Parameter estimation: ridge regression}
\label{sec:parameter_estimation}
Recall from \revise{Section~\ref{sec:model:ace_framework}} that the linear ACE models are parameterized linearly as shown in \eqref{eq:lin_ace}. As described in \revise{Section~\ref{sec:data}} we estimate parameters by matching the model to observations of total energies, forces and virials evaluated via a high fidelity reference model on different training structures $R \in {\bm R}$, where ${\bm R}$ denotes the training set. To estimate the parameters we minimize the loss function \eqref{eq:lsq-loss}. In the current section, we go into further details of the parameter estimation process once the model and training set have been specified.
 
First, we discuss the regression weights $w_{E,R}$, $w_{F,R}$ and $w_{V,R}$, which allow users to specify the relative importance of different observations and structures. In principle one could specify individual weights for each structure $R$ and observation type $E, F, V$. In practice, it has proven convenient to label all structures $R$ with a {\it configuration type} as described in \revise{Section~\ref{sec:data}} and to assign weights according to such groups. In addition the weights $w_{E,R}, w_{V,R}$ should scale like $1/\sqrt{N_R}$ where $N_R$ denotes the number of atoms in the structure $R$~\cite{bartok2010gaussian, performant2022lysogorskiy}. Thus, the weights $w_{E,R}, w_{V,R}$ take the form 
\[
    w_{E,R} = \frac{\tilde{w}_{E,{\rm cfgtype}(R)}}{\sqrt{N_R}}, \qquad 
    w_{F,R} = \tilde{w}_{F, {\rm cfgtype(R)}}, \qquad 
    w_{V,R} = \frac{\tilde{w}_{V,{\rm cfgtype}(R)}}{\sqrt{N_R}},
\]
with $\tilde{w}_{*,{\rm cfgtype}}$ defined by the user as follows: Suppose, for example, that a training set contains several solid phase structures as well as liquid structures, then we may wish to demand a higher fit accuracy on the solid structures. In addition we typically find that energies must be given higher weights in order to achieve the best possible balance of accuracy. This might result in weight specifications as shown in Listing~\ref{code:acefitscript}, lines 4-5.

\begin{listing}[H]
\begin{mdframed}
\begin{minted}[breaklines, escapeinside=||, xleftmargin=30pt, linenos=true]{julia}
model = ... # specify a model; see e.g. Listing |\ref{code:acemodel}| 
data = ...  # load training data; see e.g. Listing |\ref{code:readextxyz}|
P = smoothness_prior(model)        # regularisation operator; see |\S \ref{sec:basis_selection}|
weights = Dict( "default" => Dict("E" => 30.0, "F" => 1.0, "V" => 1.0),
                 "liquid" => Dict("E" =>  5.0, "F" => 0.5, "V" => 0.25) )
solver = BLR(tol = 1e-3, P = P)    # specify the solver, see Table |\ref{tbl:solvers}| for options
acefit!(model, data, solver; weights=weights)    # solve lsq problem, update model parameters

# model accuracy on a test set 
testdata = ... # load test data 
errors(testdata, model) 

# export the fitted potential 
export2json("model.json", model)
export2lammps("model.yace", model)
\end{minted}
\end{mdframed}
\caption{Prototypical parameter estimation script, using some simple control over regression weights and solver parameters.   }
\label{code:acefitscript}
\end{listing}

Next we discuss the minimization of the loss. Since all observations we consider here are linear, the minimization of $L(\cc)$ can be rewritten in the form 
\begin{equation} \label{eq:lsq-unregularised}
      \underset{\mathbf{c}}{\arg\min}  \, \, \big\| \mathbf{W}({\bf y} - {\bf A} \cc)  \big\|^{2},
\end{equation}
where ${\bf y}$ is a vector containing the observation values $\mathcal{E}_R, \mathcal{F}_R, \mathcal{V}_R$, ${\bf A}$ is the design matrix containing the ACE basis values corresponding to those observations and ${\bf W}$ a diagonal matrix containing the weights $w_{E,R}, w_{F,R}, w_{V,R}$. Solving the linear least squares system \eqref{eq:lsq-unregularised} often results in overfitting, hence one almost always employs regularized methods, for example the ridge regression formulation, 
\begin{equation} \label{eq:lsq-regularised}
      \underset{\mathbf{c}}{\arg\min}  \, \, \big\| \mathbf{W}({\bf y} - {\bf A} \cc)  \big\|^{2}
                + \lambda \big\| {\bm \Gamma} \cc \big\|^2,
\end{equation}
where ${\bm \Gamma}$ specifies the form of the regularizer and $\lambda$ a scaling parameter determining the relative weight of the regularisation. This formulation of the least squares problem is often also called regularized least squares, and the $\lambda \|\Gamma \cc\|^2$ term is often called generalized \revise{Tikhonov} regularisation. The default for ${\bm \Gamma}$ is zero or the identity, depending on the choice of solver. Our recommendation is to use the smoothness prior introduced in \eqref{eq:smoothness_prior} instead for most solvers. Automatic relevance determination (ARD) is unique amongst the ridge regression solvers available in {\tt \revise{ACEpotentials}.jl} in that it estimates a regularizer ${\bm \Gamma}$ from the sensitivity of the parameters to the training data, at additional computational cost; see \revise{Section~\ref{sec:bayes}} for more details.  

To solve the ridge regression problem \eqref{eq:lsq-regularised}, {\tt \revise{ACEpotentials}.jl} employes the package {\tt ACEfit.jl}\footnote{\tt https://github.com/ACEsuit/ACEfit.jl}, which offers a range of such algorithms. In the simplest setting, it can be used as shown in Listing~\ref{code:acefitscript}, lines 6-7.
For a list of the most important solvers, see Table~\ref{tbl:solvers}. For large models and/or large datasets, the parameter estimation task can be computationally challenging and may have to be performed on a cluster.

For small and moderate datasets we normally recommend the BLR method. For large datasets. when finely tuned regularisation is often less important, the random matrix sketching RRQR and iterative LSQR may be more appropriate.

\begin{table}
\begin{tabular}{r|p{16cm}}
    {\tt QR} 
         & {\bf QR decomposition:} Direct solution of the ridge regression problem \eqref{eq:lsq-regularised}. \revise{Tikhonov} regularisation is imposed by extending the linear system. This method should rarely be used in practice and is included mostly for theoretical interest and the sake of completeness. \\[1mm] 
         & {\tt solver = QR(lambda = 0.0)} \\[2mm]
    {\tt LSQR}
        & {\bf Krylov method:} the standard iterative Krylov algorithm to solve the ridge regression problem \eqref{eq:lsq-regularised}. \revise{Tikhonov} regularisation is imposed implicitly in the algorithm, with {\tt damp} corresponding to the parameter $\lambda$. Early termination, by adjusting {\tt atol} provides an additional and different form of regularisation. This algorithm is suitable for very large-scale parameter estimation problems. \\[1mm]
        & {\tt solver = LSQR(damp = 1e-4, atol = 1e-6)} \\[2mm]
    {\tt RRQR} 
        & {\bf Rank-revealing QR decomposition:} A random matrix sketching approach, which is computationally more efficient than the standard QR decomposition. In addition, the parameter {\tt rtol} is closely related to $\lambda$ in \eqref{eq:lsq-regularised} but not identical. Instead of adding a \revise{Tikhonov} term, RRQR regularisation is imposed by removing highly sensitive subspaces as determined by {\tt rtol}. For large problems, this algorithm is more performant than the standard QR decomposition. \\[1mm]
        & {\tt solver = RRQR(rtol = 1e-5)} \\[2mm]
    {\tt BLR} 
        & {\bf Bayesian Linear Regression:} (or, Bayesian ridge regression) specifies a class of solvers that estimate regularisation hyperparameters, depending on the setting it estimates the scaling parameter $\lambda$ or the entire \revise{Tikhonov} matrix $\Gamma$. This solver also determines a posterior model distribution that can be used for uncertainty quantification. See \revise{Section~\ref{sec:bayes}} for further details. This algorithm is more robust than {\tt QR, LSQR, RRQR}, but computationally more intensive. It is highly recommended for relatively small datasets. \\[1mm] 
        & {\tt solver = BLR()}
\end{tabular}
\caption{Table of solvers for the ridge regression problem \eqref{eq:lsq-regularised}.}
\label{tbl:solvers}
\end{table}

Once the model parameters are determined as shown above, we typically wish to perform two tasks: (1) confirm the model accuracy on a test set; and (2) export the model to a format that can be used in standard MD codes, e.g., LAMMPS and ASE. Suppose that we are provided with a test data set {\tt testdata}, then we can determine the model errors on that test set as seen in Listing~\ref{code:acefitscript}, lines 9-11. This will print tables of RMSE and MAE errors for individual configuration types. 
If we wish to store and/or export the fitted potential for later use, we typically save it in {\tt .json} format which can be read by {\tt \revise{ACEpotentials}.jl} as well as its Python interface to ASE, and in {\tt .yace} format which can be read by the {\tt pace} extension to LAMMPS; cf. Listing~\ref{code:acefitscript}, lines 13-15.

\subsection{Bayesian framework for parameter estimation}
\label{sec:bayes}
Uncertainty estimates of model predictions are highly sought after tools to judge the accuracy of a prediction during simulation with a fitted model, but can also be employed to great effect during the model development workflow, e.g., in an active learning context.
Such uncertainty estimates can be derived in a principled way by recasting the ridge regression problem \eqref{eq:lsq-regularised} in a Bayesian framework where inference is based on the Bayesian posterior distribution
\begin{equation}\label{eq:generic:posterior}
{\rm post}(\cc) = p(\cc \given {\bf A}, {\bf y}) \propto p( {\bf A}, {\bf y} \given \cc) \, p(\cc).
\end{equation}
Here, $p( {\bf A}, {\bf y} \given \cc)$ denotes the likelihood of the observed data, and $p(\cc)$ the prior distribution on the model parameters. The Bayesian analogue of \eqref{eq:lsq-regularised} is a Bayesian Linear Regression model with Gaussian observational noise and prior, 
\begin{align}
p( {\bf A}, {\bf y} \given \cc) &\propto \exp \left (- \frac{1}{2}({\bf y} - {\bf A} \cc)^T \revisetwo{(\beta{\bf W}^2)}({\bf y} - {\bf A} \cc)\right ), \qquad \text{and} \\ 
p(\cc)  &\propto \exp\left (- \frac{1}{2} \cc^T \mathbf{\Sigma}^{-1}_0 \cc  \right),
\end{align} 
\revisetwo{
where the covariance $\beta^{-1}{\bf W}^{-2}$ of the observation noise depends on the regression weight matrix ${\bf W}$ and a hyper-parameter $\beta>0$. This choice of prior and noise model yields a Gaussian posterior distribution,} $p(\cc \given {\bf A}, {\bf y}) = \mathcal{N}(\cc;\,\boldsymbol{\mu},\mathbf{\Sigma})$, with mean and covariance given, respectively, by $\boldsymbol{\mu} = \revisetwo{\beta\mathbf{\Sigma}{\bf A}^T{\bf W}^2 {\bf y}}$ and $\mathbf{\Sigma} = \revisetwo{ \left(\beta{\bf A}^T{\bf W}^2{\bf A} + \mathbf{\Sigma}^{-1}_0\right)^{-1}}.$
\revisetwo{
We assume that the prior covariance $\mathbf{\Sigma}_0$ is of the form of a diagonal matrix. The above Bayesian model can be connected to the ridge regression formulation of equation \eqref{eq:lsq-regularised} by noticing that maximising the posterior density \eqref{eq:generic:posterior} is equivalent to minimizing the regularized loss in \eqref{eq:lsq-regularised} when $\mathbf{\Sigma}^{-1}_0 = \zeta\mathbf{\Gamma}^2$ for some $\zeta>0$ and $\lambda={\zeta}/{\beta}$. 
}
\subsubsection*{Solvers and model selection via evidence maximisation}
The reliability of uncertainty estimates critically depends on the values of the model hyper-parameters, the noise and prior covariance matrices $\revisetwo{\beta^{-1}{\bf W}^{-2}}$ and $\mathbf{\Sigma}_0$. In ACE, it is sometimes difficult to make informed guesses of explicit values of these hyper-parameters that lead to good fits. We therefore commonly employ empirical Bayes approaches that infer appropriate values of these parameters directly from the training data by virtue of maximising the model evidence 
\begin{equation}
\begin{aligned}
     p({\bf A}, {\bf y} \given \mathbf{\Sigma}_0, \revisetwo{\beta})& = \int p({\bf A}, {\bf y} \given \cc,  \revisetwo{\beta}) p(\cc \given \mathbf{\Sigma}_0) d\cc\\
     &=     \sqrt{\frac{\revisetwo{\beta}(2\pi)^{-N_{\rm obs}}|\mathbf{\Sigma}|} { |\mathbf{\Sigma}_0| |\revisetwo{{\bf W}^{-2}}| } }
\exp \left (- \frac{1}{2}  (\mathbf{y} - \mathbf{A} \boldsymbol{\mu})^T  \revisetwo{(\beta{\bf W}^2)} (\mathbf{y} - \mathbf{A} \boldsymbol{\mu}) - \frac{1}{2}  \boldsymbol{\mu}^T\mathbf{\Sigma}^{-1}_0 \boldsymbol{\mu}  \right)
\end{aligned}
\end{equation}
as a function of $\mathbf{\Sigma}_0, \revisetwo{\beta}$.
Intuitively, maximising the model evidence results in a model where the regularising effect of the covariance matrix $\mathbf{\Sigma}_{0}$ and the degree of penalisation of model misfit---modelled by the noise covariance matrix $\revisetwo{\beta^{-1}{\bf W}^{-2}}$---are balanced against the degree to which the regression coefficients are determined by the data. 

Within {\tt \revise{ACEpotentials}.jl} this is implemented in the {\tt BLR} solver (cf. Table~\ref{tbl:solvers}). Different solver options result in different constraints on \revisetwo{the form of the prior covariance} $\mathbf{\Sigma}_0$, and we refer to the documentation~\cite{ACEpotentials} for further details.

\subsubsection*{Uncertainty estimates via committees}
\label{sec:uncertainty}

Formally, the Bayesian ridge solver provides not an optimal parameter vector $\cc$ but a posterior parameter distribution $p(\cc)$. 
In practice, one then selects the mean parameter  vector ${\bm \mu}$ to specify the model. 
However, the posterior distribution remains important to estimate the uncertainty of predictions. 
Evaluating such uncertainties from the exact posterior distribution is computationally expensive; instead, {\tt \revise{ACEpotentials}} draws $K$ samples $\{ \cc_{k}  \}_{k = 1}^{K}$  from ${\rm post}(\cc)$ resulting in a committee of ACE models which can be used to obtain computationally efficient uncertainty estimates for predictions. 
For example, the standard deviation $\sigma$ of a total energy prediction can be approximated by a committee via 
\begin{equation}
\begin{aligned}
\tilde{\sigma}^{2} = \revisetwo{\frac{1}{\beta w_{E,R}^2}} + \frac{1}{K}\sum_{k=1}^{K} (E^{k} - E^{\boldsymbol{\mu}})^{2},
\end{aligned}
\label{eq:e_uncertainty}
\end{equation}
where $E^{\boldsymbol{\mu}}$ is the prediction made by the mean model with parameters $\boldsymbol{\mu}$, while $E^{k}$ are the committee predictions from models with parameters $\cc_k$. 
Similarly, uncertainty estimates can be made for any partial derivative of the potential energy surface such as for committee forces $F^{k} = \cc_k \cdot \nabla \BB_i$, or the mean force $F^{\boldsymbol{\mu}} = \boldsymbol{\mu} \cdot \nabla \BB_i$. 

The first term in \eqref{eq:e_uncertainty} refers to the aleatoric, or irreducible, uncertainty arising due to randomness of the system which is dominated by the complexity of the linear ACE convergence parameters such as correlation order, polynomial degree and cutoff.
The second term is the epistemic, or reducible, uncertainty arising due to a lack of data or rather information. 
An example how a variance estimate of the epistemic uncertainty can be obtained in the linear ACE framework is shown in Listing~\ref{code:committee}.

\begin{listing}[H]
\begin{mdframed}
\begin{minted}[breaklines, escapeinside=||, xleftmargin=30pt, linenos=true]{julia}
E, E_co = co_energy(model.potential, atoms)
sigma = sqrt( mean( (E_co .- E).^2 ) )
\end{minted}
\end{mdframed}
\caption{Example how to use a committee to estimate the uncertainty of a prediction. \revise{(Note that {\tt model.potential} gives access to the calculator object.)} Analogously, one can obtain committees of forces and virials.}
\label{code:committee}
\end{listing}

\section{Workflow Examples}
\label{sec:workflow}

\revise{In this section, we present several practical examples of ACE usage, including simple benchmarks, practical potentials for materials and liquids to examples illustrating the hyperactive learning workflow. The scripts we used to generate the reported results are made available in a separate git repository\footnote{\tt https://github.com/ACEsuit/ACEworkflows} that will be regularly updated as the {\tt ACEpotentials.jl} package evolves.} 

\subsection{Tests with pre-existing data sets}

\label{sec:pre-existing_data_sets}

\subsubsection{Benchmarks with limited-diversity datasets}
We test {\tt \revise{ACEpotentials}.jl} with default parameters on an early single-element benchmark dataset taken from \cite{Zuo_2020}. This dataset was originally used to assess the relative strengths and weaknesses of four important MLIPs, the high-dimensional neural network potential (NNP)\cite{NNP2007}, the Gaussian approximation potential (GAP) \cite{GAP2010}, the Spectral Neighbor Analysis Potential (SNAP)\cite{THOMPSON2015SNAP}, and moment tensor potentials (MTP)\cite{MTP2016}. The benchmark contains six separate datasets corresponding to the six elements Li, Mo, Ni, Cu, Si and Ge, spanning a variety of chemistries (main group metal, transition metal and semiconductor), crystal structures (bcc, fcc, and diamond) and bonding types (metallic and covalent).
For each element, the dataset contains the ground-state crystal structure, strained structures with strains of -10\% to 10\%, slab structures up to a maximum Miller index of three, and NVT ab initio molecular dynamics simulations of the bulk supercells with and without a single vacancy. These datasets contain a relatively large number of training structures, but only limited diversity.

In table \ref{limited diversity dataset} we see the comparison of the MAEs in energies and forces for the best performing potentials in the benchmark (GAP and MTP) with two linear ACE models trained with the default parameters and total degrees chosen to reach basis sizes of, respectively, 300 basis functions for {\bf ACE(s)} and approximately 1000 basis functions for {\bf ACE(l)}. We optimized none of the hyperparameters and solved used RRQR to estimate the parameters. We chose RRQR since the datasets are very large, hence a highly tuned regularisation is less important.
This results in competitive accuracy across the entire benchmark. The only small exception is the slightly larger energy error for Mo-ACE(l), which suggests some fine-tuning of the model parameters could be beneficial in this particular case.
Our aim with this experiment was to demonstrate that, with only minimal effort, linear ACE models can perform with (near-) best accuracy in a data set geared towards testing statistical generalization.

\begin{table}
    Energy [meV] \hspace{3.8cm} Forces [eV/\AA] \\[2mm] 
\begin{tabular}{rrrrr}
  \hline
   & \textbf{ACE(sm)} & \textbf{ACE(lge)} & \textbf{GAP} & \textbf{MTP} \\\hline
  Ni & 0.416 & 0.34 & 0.42 & 0.48 \\
  Cu & 0.292 & 0.228 & 0.46 & 0.41 \\
  Li & 0.231 & 0.165 & 0.49 & 0.49 \\
  Mo & 2.597 & 2.911 & 2.24 & 2.83 \\
  Si & 3.501 & 1.985 & 2.91 & 2.21 \\
  Ge & 2.594 & 2.162 & 2.06 & 1.79 \\\hline
\end{tabular}
\begin{tabular}{rrrrr}
  \hline
   & \textbf{ACE(sm)} & \textbf{ACE(lge)} & \textbf{GAP} & \textbf{MTP} \\\hline
  Ni & 0.018 & 0.015 & 0.02 & 0.01 \\
  Cu & 0.007 & 0.005 & 0.01 & 0.01 \\
  Li & 0.006 & 0.005 & 0.01 & 0.01 \\
  Mo & 0.123 & 0.097 & 0.09 & 0.09 \\
  Si & 0.086 & 0.066 & 0.07 & 0.06 \\
  Ge & 0.064 & 0.051 & 0.05 & 0.05 \\\hline
\end{tabular}    

  \caption{Mean absolute test errors in predicted energies and forces of two ACE models, ACE(sm) with ca 300 basis functions and ACE(lge) with ca 1000 basis functions, compared against the two best performing MLIPs published in \cite{Zuo_2020}. }
  \label{limited diversity dataset}
\end{table}

\subsubsection{Silicon}
\begin{figure}[hb]
\centering
\begin{minipage}[b]{\textwidth}
\includegraphics[width=0.7\columnwidth]{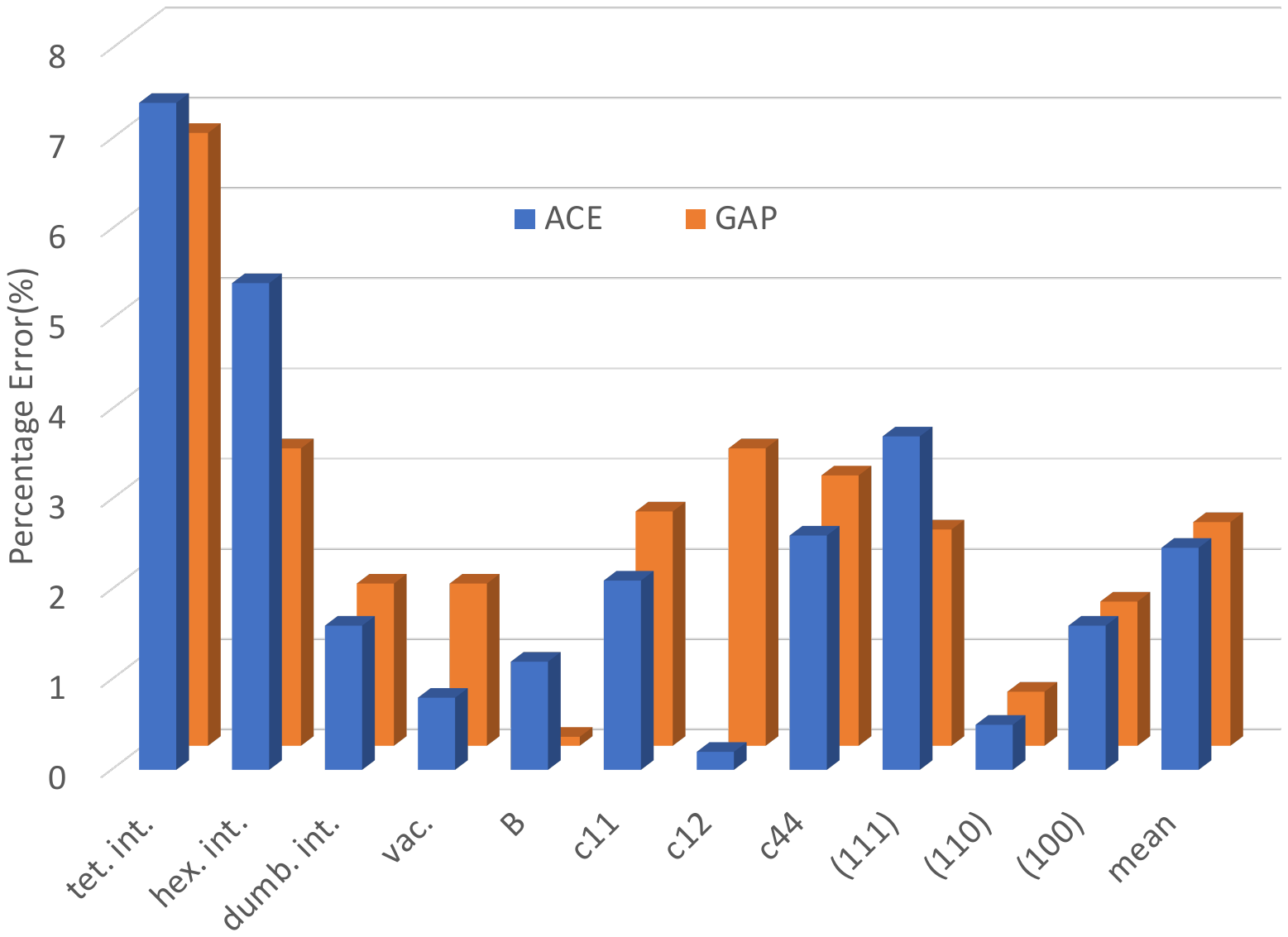}
\begin{tabular}{c|cccc|cccc|ccc}
 & \multicolumn{4}{c|}{Point defects [eV]} & \multicolumn{4}{c|}{Elastic Properties [GPa]} & \multicolumn{3}{c}{Surface energy [J/m$^2$]}\\
 & tet. int. & hex. int. & dumb. int. & vac. & $B$ & $c_{11}$ & $c_{12}$ & $c_{44}$ & (111) & (110) & (100) \\
\hline
 DFT & 3.91 & 3.72 & 3.66 & 3.67 & 88.6 & 153.3 & 56.3 & 72.2 & 1.57 & 1.52 & 2.17   \\
  \multicolumn{5}{c}{} & \multicolumn{4}{c}{Relative Error [$\%$]} \\
\hline
 GAP & 6.8 & 3.3 & 1.8 & 1.8 & 0.1 & 2.6 & 3.3 & 3 & 2.4 & 0.6 & 1.6\\
 ACE & 7.4 & 5.4 & 1.6 &	0.8 & 1.2 &	2.1 & 0.2 &	2.6 & 3.7 &	0.5 & 1.6 \\ 
\hline
\end{tabular}
\end{minipage}
\caption{Benchmark of the silicon GAP \cite{Bartok2018-fk} and ACE model presented in this work. Percentage relative errors against the DFT reference are provided in the Table }
\label{fig:si_properties}
\end{figure}

We used {\tt \revise{ACEpotentials}.jl} to fit a linear ACE potential to the silicon dataset introduced by Bart{\'o}k et al~\cite{Bartok2018-fk} for fitting a Gaussian approximation potential (GAP). 
This extensive database contains a wide range of configurations ranging from several bulk crystal structures (diamond, hcp, fcc, etc.), amorphous structures as well as liquid MD snapshots, aiming to cover as much of the silicon energy landscape as possible.
The corresponding GAP model was shown to outperform a wide range of other (classical) interatomic potentials on a large selection of accuracy and property or generalisation tests ranging from surface formation energies as well as liquid and radial distribution functions.
The current work benchmarks \revise{an} {\tt \revise{ACEpotentials}.jl} model, with default model parameters, containing basis functions up to order $\bar{\nu}=4$, polynomial total degree $D^{\max}=\revise{20}$ and \revise{6}~\AA\ cutoff against this silicon GAP potential.
\revise{
The model was fitted using generalised Tikhonov regularisation \eqref{eq:lsq-regularised} of $\lambda \Gamma$, where $\Gamma$ was constructed using an algebraic smoothness prior \eqref{eq:smoothness_prior} with $p=5$, whilst the \texttt{BLR} solver was used to estimate the scaling parameter $\lambda$.
}
This benchmark is formed of a series of property tests including bulk diamond elastic constants, vacancy formation energies, surface formation energies for the (100), (110), (111) surfaces and hexagonal, dumbbell and tetragonal point defect energies for bulk diamond.
These results of these property tests for the CASTEP \cite{ClarkSegallPickardHasnipProbertRefsonPayne+2005+567+570} DFT reference, GAP and ACE are shown in Figure~\ref{fig:si_properties} and indicate good accuracy across the range of property tests. 
Percentage errors relative to the DFT reference are \revise{also included}, confirming similarly accurate performance between the GAP and the {\tt \revise{ACEpotentials}.jl} frameworks.

\revise{We also used this silicon ACE potential to carry out a more challenging test, namely to simulate fracture in the $(111)[1\bar{1}0]$ cleavage system. We used the \texttt{matscipy} package to setup a $12\times11\times1$  supercell containing 1586 atoms and to carry out structural optimisations with a Mode I crack anisotropic continuum linear elastic displacement field~\cite{Sih1965} applied with stress intensity factors ranging from 0.6$K_G$ to 1.5$K_G$ (where $K_G$ is the Griffith load at which fracture becomes thermodynamically favourable). We observed spontaneous formation of the Pandey $2\times1$ reconstructed $(111)$ surface behind the crack tip, in good agreement with previous studies using DFT~\cite{Kermode2008} and GAP~\cite{Bartok2018-fk}. The critical stress intensity factor was determined to be $K_I=1.0\pm0.02K_G$, which is very close to the expected Griffith value, indicating minimal lattice trapping. Overestimating the extent of lattice trapping is a common failure mode of previous interatomic potentials when applied to model fracture~\cite{Bitzek2015}. The total simulation time was around 30 minutes on a 28-core workstation.}

\begin{figure}[h]
    \centering
    \includegraphics[width=0.8\columnwidth]{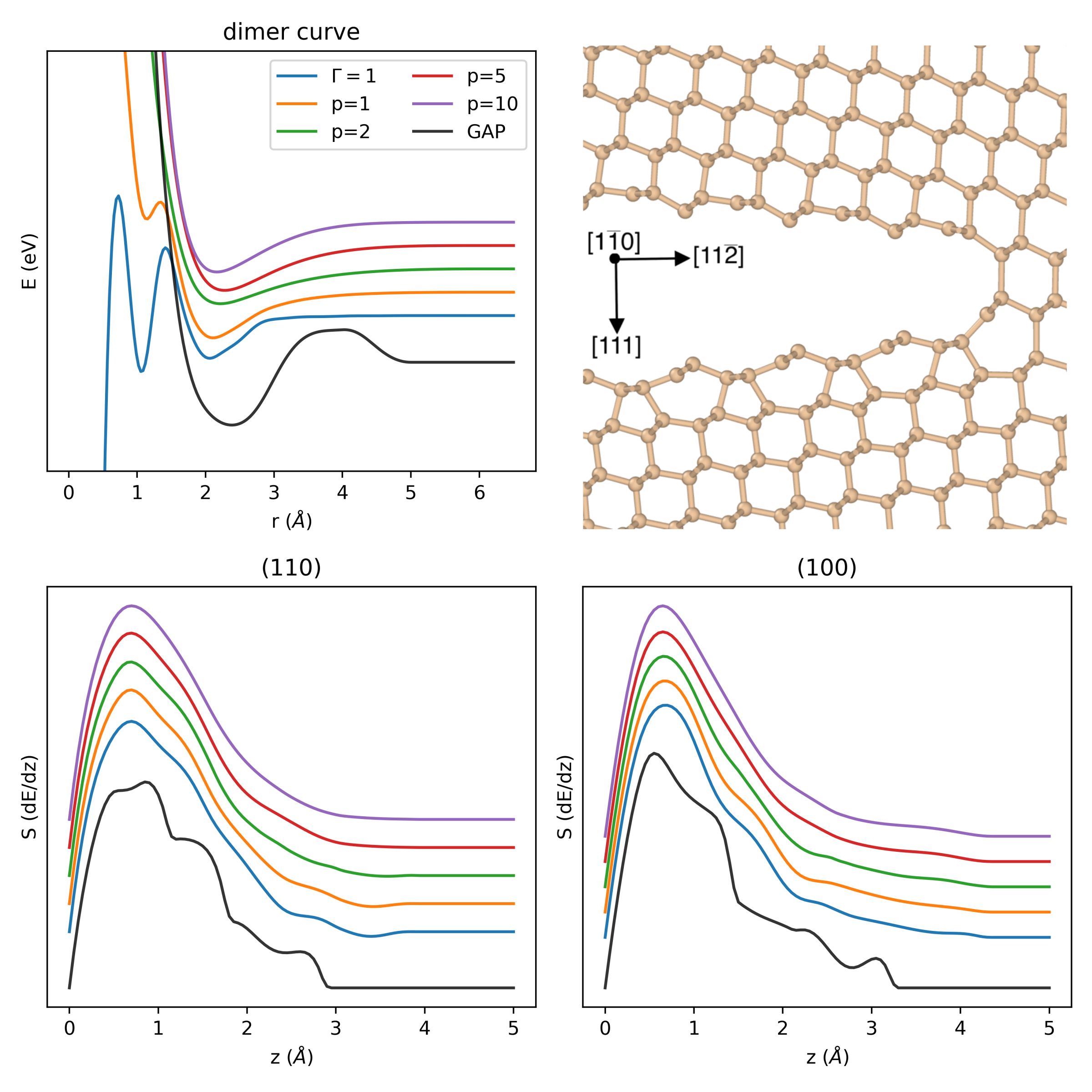}
    \caption{\revise{\textbf{Top Left}: The predicted energy of the Si-Si dimer is shown for a sequence of ACE potentials trained with varying strengths of smoothness prior but equal accuracy (Force RMSE $\approx$ 0.075 eV/\AA). $\Gamma=1$ corresponds to an equal prior for all basis functions whilst $p$ indicates the strength of the algebraic smoothness prior defined in \eqref{eq:smoothness_prior}. The black curve shows the corresponding result using GAP. All curves are shifted for clarity. \textbf{Bottom}: The evolution of stress ($S$) as a function of separation ($z$) during rigid decohesion of bulk silicon into the unrelaxed (110) and (100) surfaces is shown for the same sequence of potentials. \textbf{Top Right}: Snapshot from Si$(111)[1\bar{1}0]$ quasi-static fracture simulation at a stress intensity factor of 1.8$K_G$ using our ACE potential. The lower fracture surface shows a $2\times1$ Pandey reconstruction (alternating pentagons and heptagons), consistent with previous studies using DFT and GAP models, but at much reduced cost. The critical fracture toughness is very close to $K_G$, showing minimal lattice trapping.}}
    \label{fig:Si_decohsion_fracture}
\end{figure}

\revise{
To successfully carry out the fracture test it was crucial to produce a highly regular (smooth) ACE potential. To illustrate the effect of changing the smoothness prior, a sequence of ACE potentials (order $\bar{\nu}=4$, total degree $D^{\max}=21$ and 6~\AA\ cutoff), was fitted using no smoothness prior ($\Gamma=1$) and increasing strengths of algebraic smoothness prior \eqref{eq:smoothness_prior}, $p = 1, 2, 5$ and $10$. In all cases the model parameters were estimated using generalized Tychonov regularisation \eqref{eq:lsq-regularised} with the scale factor $\lambda$ tuned such that all potentials achieved a force RMSE of approximately $0.075$~eV/\AA\ , which is approximately $5\%$ larger than without any regularisation. The effect of the prior on predicted Si-Si dimer curves and rigid bulk Si decohesion curves, which respectively probe smoothness of 2-body and many-body terms, is shown in Figure \ref{fig:Si_decohsion_fracture}. Applying a moderate smoothness prior aids extrapolation into the close-approach region and reduces the amplitude of spurious oscillations seen in the stress (S) during decohesion.}

\revise{\subsubsection{Water}}
\revise{We investigated the ability of {\tt \revise{ACEpotentials}.jl} to capture the interactions in complex molecular liquids and to perform robust  molecular dynamics simulations in such systems, fitting a linear ACE potential to a dataset containing 1593 liquid water configurations~\cite{cheng2019ab}. 
We chose only default model parameters, containing basis functions up to correlation order $\numax = 3$, polynomial total degree $D^{\max}=15$ and $r_{\rm cut} = 5.5$~\AA\ cutoff.
Parameter estimation was performed using ARD with relevance threshold set by minmising the Bayesian Information Criterion (BIC)~\cite{10.1214/aos/1176344136}. 
The training RMSE were 1.732 meV/atom for energies and 0.099 eV/\AA~for forces.
To investigate the performance and robustness of the fitted ACE model, a series of mean squared displacement (MSD) simulation were performed under 1 bar NPT conditions at 300 K. 
The simulations were performed using 5184 atom simulation boxes, shown in Fig.~\ref{fig:water_msd} below, with PACE-LAMMPs~\cite{performant2022lysogorskiy}.
The total simulation time for each of these simulation was 20 minutes utilising 1280 cores on ARCHER2, illustrating the efficiency of ACE potentials.
The diffusion constant predicted by this simulation was 1.20 $\pm$ 0.03 m/s$^{2}$.
It should be noted that diffusion constants are notoriously difficult to accurately determine especially considering the absence of long-range interactions into these ACE models. This example is therefore mostly an illustration of robustness and performance.}

\begin{figure}[ht]
    \includegraphics[width=0.6\columnwidth]{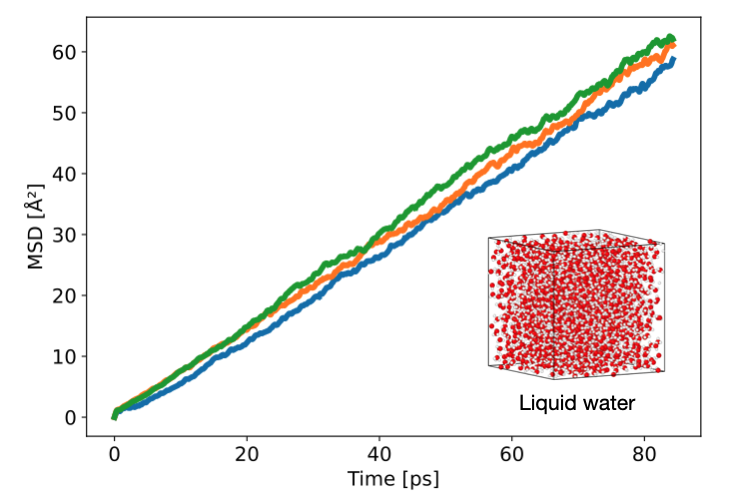}
    \caption{\revise{Mean squared displacement (MSD) for three liquid water simulation at 1 bar NPT simulations and 300 K. The simulation cell contained 5184 atoms.}}
    \label{fig:water_msd}
\end{figure}

\subsection{The Hyperactive Learning (HAL) Workflow}
While fitting ACE potentials to pre-existing or ``manually'' assembled datasets, as discussed in Section~\ref{sec:pre-existing_data_sets}, the real benefit of the linear ACE framework is in the construction robust and computationally inexpensive ACE potentials from the ground up with automated dataset assembly. This is achieved through the use of an iterative loop employing an active learning (AL) type approach \cite{vandermause2022, PODRYABINKIN2017171}, where relevant training configurations are sampled to form a training database. To accelerate this AL process, hyperactive learning (HAL) \cite{hyperactive2022} we introduced, which adds a biasing term to a molecular dynamics simulation towards predicted high uncertainty $\sigma$, as shown in~\eqref{eq:ehal}. A tunable parameter $\tau$ controls the strength of the biasing and thus the balance between physical exploration (molecular dynamics) and discovery of new structures (biasing).

\begin{equation}
    E^{\textrm{HAL}} = E^{\rm ACE} - \tau \sigma.
    \label{eq:ehal}
\end{equation}

The HAL framework shares similarities with Bayesian Optimization (BO) as the biasing term is formally equivalent to a Lower Confidence Bound (LCB) acquisition function \cite{PhysRevB.105.245404}. 
Similarly to BO, the parameter $\tau$ adjusts the tradeoff between exploration and exploitation during the generation of training configurations using HAL. 
HAL-generated configurations are both energetically reasonable, guided by $E^{\rm ACE}$ (exploitation), and informative, predicted by a relatively large value of $\sigma$ (exploration).
The bias towards uncertainty, mediated by an emerging biasing force during HAL dynamics, can be viewed as a strategy to acquire information (gain) by seeking out unseen (local) environments. 
The HAL approach can also be viewed as an adversarial attack, aimed to destabilize a fitted ACE potential such that, after iteratively adding sufficiently many new configurations, the linear ACE model is robust to such attacks which all but guarantees stable dynamics over long timescales.

The biasing parameter $\tau$ in HAL necessitates careful tuning, which HAL achieves through an adaptive scheme \cite{hyperactive2022} that tunes $\tau$ on the fly by balancing the magnitude of the biasing force relative to the forces obtained by $E^{\rm ACE}$. 
The {\em relative biasing parameter} $\tau_r$ used in this scheme is typically set to 0.1 to 0.2 and ensures that the biasing strength is reduced or increased depending on the degree of predicted uncertainty explored during the dynamics.

\begin{figure}[ht]
    \includegraphics[width=0.5\columnwidth]{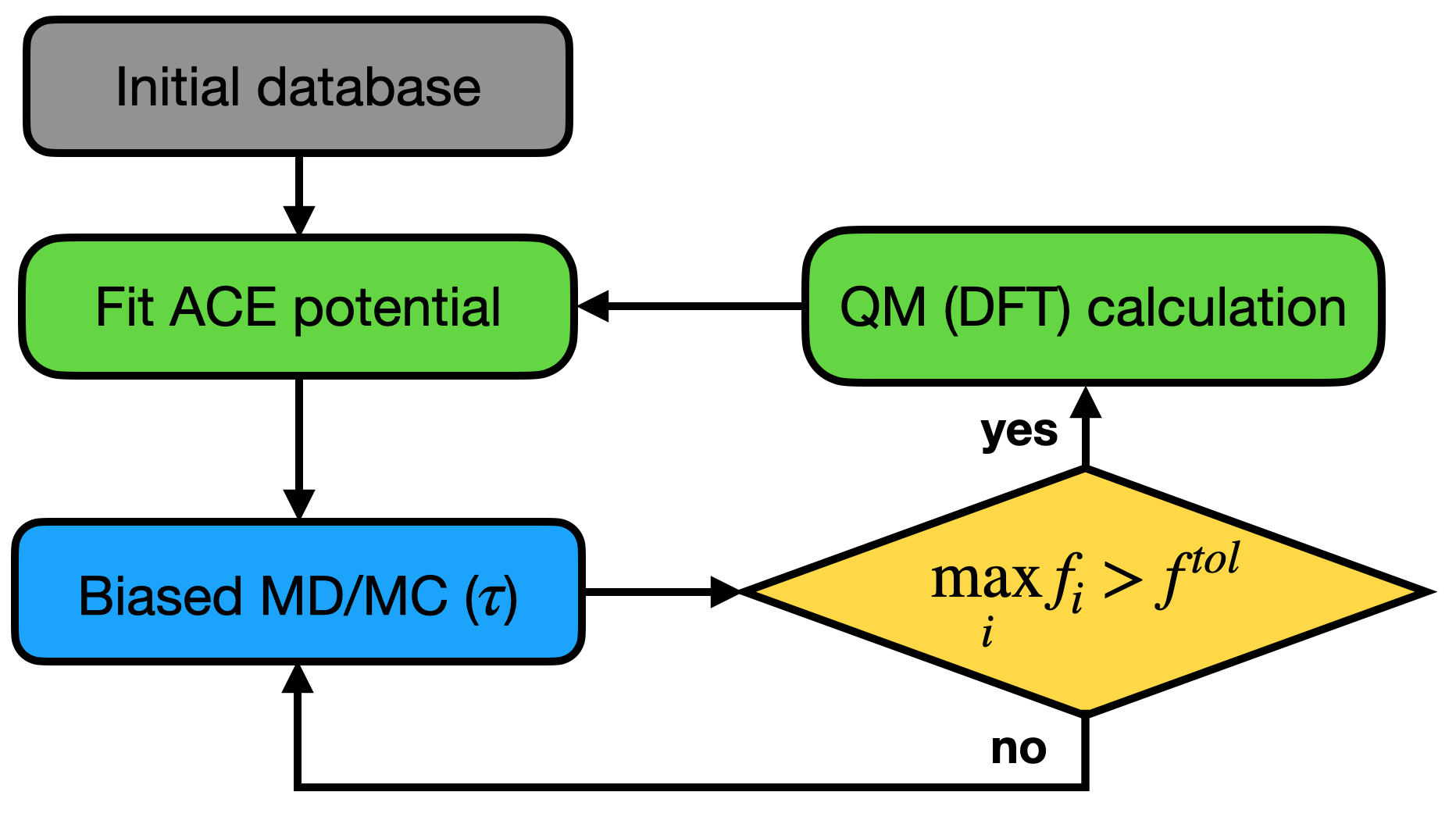}
    \caption{Hyperactive Learning (HAL) protocol. Linear ACE potentials are fitted using BRR or ARD after which biased MD/MC steps are performed controlled by biasing parameter $\tau$. Once the uncertainty metric $f_{i}$ exceeds $f^{\textrm{tol}}$ a DFT calculation is triggered \revise{a HAL iteration is completed} and the training database extended. }
    \label{fig:HAL_scheme}
\end{figure}

To initiate HAL, an initial database is typically constructed consisting of 1-10 configurations that sketch out some aspects of the energy landscape that are of interest to the application at hand. 
An ACE potential is fitted using a variant of the {\tt BLR} solver, after which committee parameterisations $\{ {\bf c}_{k} \}_{k=1}^{K}$, typically $K = 8$, are sampled from the posterior as discussed in Section~\ref{sec:bayes}. 
Biased MD/MC dynamics are then performed on $E^{\textrm{HAL}}$, using the dynamically tuned $\tau$ parameter. 
During the dynamics the relative force uncertainty $f_{i}$ is recorded and once it exceeds a predefined tolerance $f^{\text{tol}}$ a DFT calculation is triggered, and the training database is extended. 
This relative force uncertainty $f_{i}$ is defined as  
\begin{equation}
    f_{i} = \frac{\frac{1}{K} \sum_{k=1}^{K}  \| F_{i}^{k} - F^{\boldsymbol{\mu}}_{i}  \|}{\|  F^{\boldsymbol{\mu}}_{i}  \| + \varepsilon },
    \label{eq:rel_force_err}
\end{equation}
where $F_{i}^{k}$ are the forces as obtained by the committee and $F^{\boldsymbol{\mu}}_{i}$ the forces predicted by the mean $\boldsymbol{\mu}$ of the posterior over the coefficients as outlined in Sec.~\ref{sec:uncertainty}. $\varepsilon$ is a regularising constant used to regularize the fraction typically set to 0.2-0.4 eV/\AA.
Careful tuning of $f_{\textrm{tol}}$ is required as it tunes the degree of extrapolation when adding new (unseen) configurations to the training database. 
Too large $f_{\textrm{tol}}$ may lead to the sampling of energetically unreasonable configurations, whereas too small $f_{\textrm{tol}}$ leads to suboptimal information gain during the HAL scheme resulting in sampling unnecessarily many configurations. 
The HAL scheme is outlined in Figure~\ref{fig:HAL_scheme} illustrating how from a small initial training database containing a handful of configurations of interest a stable ACE potential is generated by performing biased MD and MC steps and iteratively triggering DFT calculations. \revise{For future reference, we define a {\em HAL iteration} to consist of (i) a biased MD simulation run until a new unseen structure is flagged, (ii) evaluating energies, forces and virials on the new structure, and (iii) updating the ACE potential model.}

\subsubsection{AlSi10 melting temperature}

The HAL framework was used to create an ACE potential for determining the melting temperature of the AlSi10 alloy. 
An initial dataset consisted of 32-atom random fcc lattice configurations, each containing 98 aluminium and 10 silicon atoms.
This initial dataset was composed of 5 fcc random alloy configurations with lattice constants ranging from 14.3 to 16.6~$\text{\AA}^{3}$/atom. 
The ACE basis set included interactions up to correlation order $\bar{\nu}=2$ (3-body), and employed a cutoff of 5.5~\AA. 
\revise{The model was fitted using Automatic Relevance Determination (ARD) and its sparsity set by minimising BIC which resulted in increasingly complex ACE models as more configurations (or information) were .}
The chosen maximum polynomial degree $D^{\max}$ during the HAL procedure increased from 4 to 12. 
The parameter estimation was carried out using ARD. The HAL relative  biasing strength was set to $\tau_{r}=0.2$, and the relative uncertainty threshold to $f^{\textrm{tol}}=0.2$. 

The HAL dynamics was used to melt the random alloy crystal structure, by ramping the temperature from 0~K to 1500~K at 1~GPa using a 1~fs timestep. 
Cell swapping and volume adjusting HAL-MC steps were taken to facilitate exploration of the (biased) energy landscape. After 18 HAL iterations, the ACE potential was already able to consistently perform 5000 HAL MD/MC timesteps without encountering new structures with high uncertainty. 
This final ACE potential contained 79 basis functions as selected using ARD pruning. 

\begin{figure}[ht]
    \includegraphics[width=0.8\columnwidth]{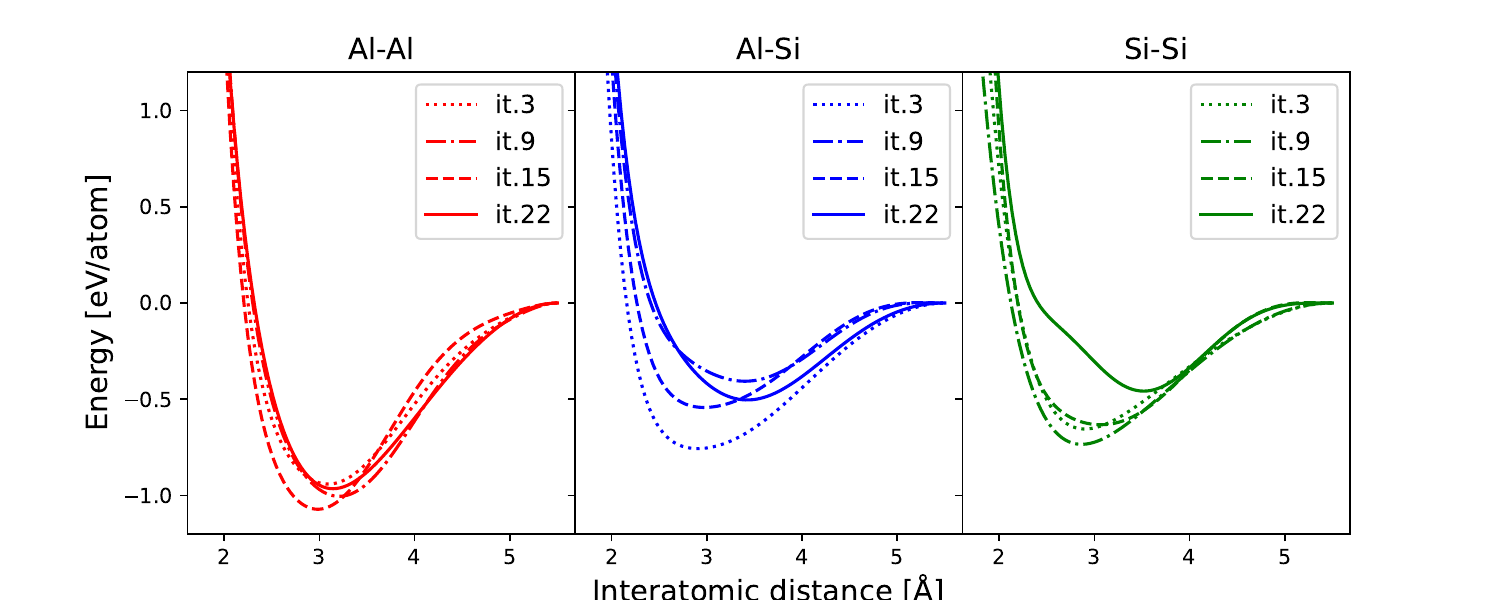}
    \caption{ACE dimer curves for pair interactions for several HAL iterations. Stronger colours indicate later HAL iterations. They key observation to be drawn from this figure is that even in the early stages of the HAL process with very little available data, our priors ensure that the dimer curves are physically sensible, in particular smooth and repulsive.}
    \label{fig:AlSi10_dimers}
\end{figure}

During these 18 HAL iterations the dimer curves are typically examined to ensure the potentials exhibit attraction at typical interatomic distances and short range repulsion as illustrated in Fig.~\ref{fig:AlSi10_dimers}. 

\begin{figure}[ht]
    \includegraphics[width=0.6\columnwidth]{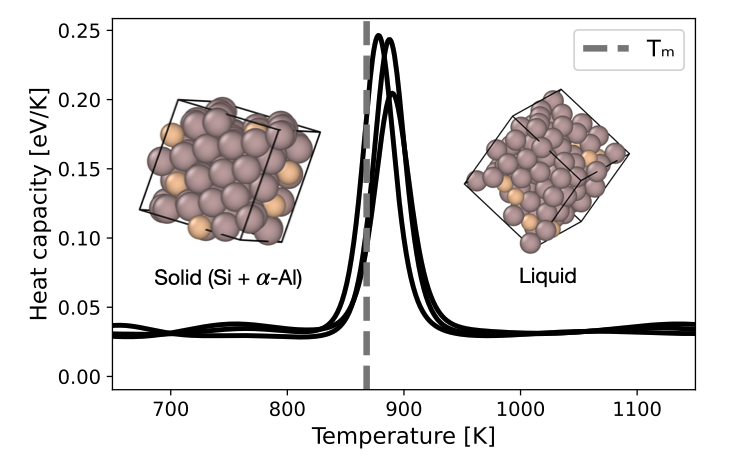}
    \caption{NS AlSi10 heat capacity curves for several runs indicating the  liquid-solid transition as predicted by the HAL generated ACE potential.}
    \label{fig:AlSi10_NS}
\end{figure}

The ACE potential obtained after HAL iteration 18 (fitted to 22 structures in total) was subsequently used to perform nested sampling (NS) simulations to model the liquid-solid phase transition. 
NS simulations were performed using 384 NS walkers, using a total decorrelation length of 512 formed by volume/shear/stretch/swap MC steps at a ratio of 4:4:4:4. 
The resulting heat capacity curves obtained by NS are presented in Figure~\ref{fig:AlSi10_NS} and are in close agreement to the melting temperature of 867~K as given by Thermo-Calc using the TCAL4 database \cite{Tang2016}.

\subsubsection{Polyethylene glycol}
\begin{figure}[ht]
    \includegraphics[width=0.6\columnwidth]{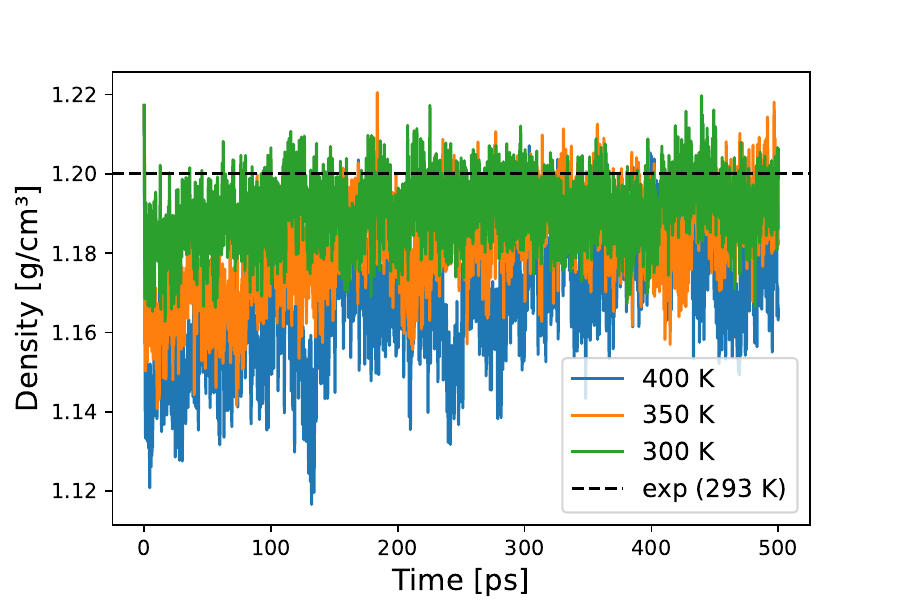}
    \caption{PEG($n$=200) density for HAL generated ACE potential under periodic boundary conditions using LAMMPs.}
    \label{fig:PEGdensity200}
\end{figure}

The HAL framework \cite{ACEHAL} was used to create a polyethylene glycol (PEG) model. 
To initilize HAL, 18 structures of PEG($n$=32) formed of 32 monomer units in vacuum were evaluated using the ORCA code~\cite{ORCA} with the $\omega$B97X DFT exchange correlation functional~\cite{ORCA_ex_cor_func} and the 6-31G(d) basis set. 
ACE models were fitted to the initial and subsequent datasets with correlation order $\bar{\nu}=3$, total degree $D^{\max}=12$ and a cutoff radius 5.5~\AA, using the ARD algorithm.
The HAL protocol used relative biasing parameter $\tau_{r}=0.1$ and uncertainty tolerance $f^{\rm tol}=0.3$ and performed at 500~K.
Unlike the previous AlSi10 example, no cell adjusting or atom swapping HAL-MC steps were performed as the configurations are isolated molecules in vacuum.
It was also chosen to not change the ACE basis throughout the HAL procedure but rather to keep it constant (e.g. $D^{\max}=12$) as the initial database was relatively diverse. 
After 50 HAL iterations an ACE potential was generated that was deemed stable as it completed 10$^4$ HAL biased MD steps without triggering a DFT calculation. It was then used to determine the density of a PEG polymer formed of $n=200$ monomer units in LAMMPS under periodic boundary conditions using the PACE evaluator \cite{performant2022lysogorskiy}. 
The PEG($n$=200) density was determined at 300~K,
350~K and 400~K at 1 bar pressure over a timescale of
0.5 ns as shown in Figure~\ref{fig:PEGdensity200}.
The density at 300~K is in good agreement with the experimental density of 1.2~g/cm$^{3}$ \cite{PEG_density_reference} at 293~K. 
This illustrates remarkable extrapolative performance by the linear ACE framework as the DFT reference (ORCA) does not support periodic boundary conditions itself, making determining the PEG density purely from first principles impossible. 

\subsubsection{Perovskite CsPbBr$_3$}

We used the HAL framework~\cite{ACEHAL} to create a training dataset for the lead-halide perovskite CsPbBr$_3$, which shows 
three relevant phases: orthorhombic at low temperatures, tetragonal at intermediate temperatures, and cubic at high temperatures, with experimental transition temperatures of 361~K and 403~K~\cite{2013_stoumpos}. 
The HAL process was designed to sample all of these phases so that the resulting potential accurately represents energy and entropy of each phase and is hence capable of predicting the transition temperatures. 
To ensure consistent DFT energies and effective vibrational mode sampling, approximately cubic 40 atom supercells were created for all three phases.

This problem required some refinement of the standard HAL procedure, and careful testing of fitted ACE potentials for several basis sizes. We therefore give more detail about the process than in the previous cases.

The initial fit starting the HAL process used a set of 15 randomly perturbed (unit cell and atomic
positions) 40-atom configurations, three from each of the high symmetry phases. The default ACE basis was used, with a 
cutoff of 8~\AA, a smoothness prior with $p=3$, and the {\tt sklearn} {\tt BayesianRidge} linear solver. 
Automated basis selection was applied every 10 HAL iterations, with a maximum basis size of 2000, $\bar{\nu}=3$, a maximum total polynomial degree of 16, and the model score as the selection criterion.  
To encourage exploration of a wide range of temperatures and configurations, over a maximum of 10$^4$ 1~fs HAL MD steps the temperature was ramped from 200~K to 600~K, and $\tau_r$ from 0.1 to 0.5. 
New fitting configurations were selected when the fractional force error $f^{\text{tol}}$ exceeded 0.4. 
After 20 iterations starting from the three unperturbed 
high-symmetry 40-atom cells at fixed unit cell shape and size, the process was restarted from 9 80-atom high symmetry cells, 
doubling each of the three 40-atom cells along each cell vector, for 20 additional iterations. Then 20 additional
iterations were carried out with variable unit cell and an applied pressure of 0.

At this point the model appeared to be stable enough \revise{for 10$^5$} steps without a HAL bias, so we switched to an unbiased
sampling process to gather more data and improve the model accuracy.
Starting the fit from the complete set of configurations 
from the HAL process, we generated fitting configurations from 2000 step runs
with a maximum basis size of 4000. These used the same 80 atom starting configurations, but
at fixed temperatures of 200~K to 500~K at 100~K intervals, and fixed shape but
variable unit cell volume. To further refine the performance of the low energy parts of the PES around
each high symmetry structure, we sampled 36 more configurations, each with 160 atoms (the three 40 atom supercells 
doubled along each of the three pairs of lattice vectors) at a range of lower temperatures, 150~K to 300~K at 50~K intervals. 

The original set of 15 randomly perturbed configurations, another similar set of 15, and the 168 HAL configurations were used as
the reference database for a set of fits to explore the performance of the model for a wide range of basis sizes. At this stage we filtered out physically unreasonable
fitting data, as defined by a criterion that excluded any force larger than 10~eV/\AA, as well as 
the energies and virials from such configurations.  To fit the model and evaluate its predictive accuracy we split
the set of configurations into 75\% fitting and 25\% testing, stratifying the split by the HAL iteration (or
initial random perturbation set) that produced the configuration. The same fitting procedure and basis as in
the HAL run were used, with $\bar{\nu}=2$ and $\bar{\nu}=3$ and maximum polynomial degree 4 to 16, up to a maximum 
basis size of $2 \times 10^{4}$. We also compared three choices for the smoothness prior: none, $p=2$, and $p=4$. 

\begin{figure*}
    \centering
    \includegraphics[width=\columnwidth]{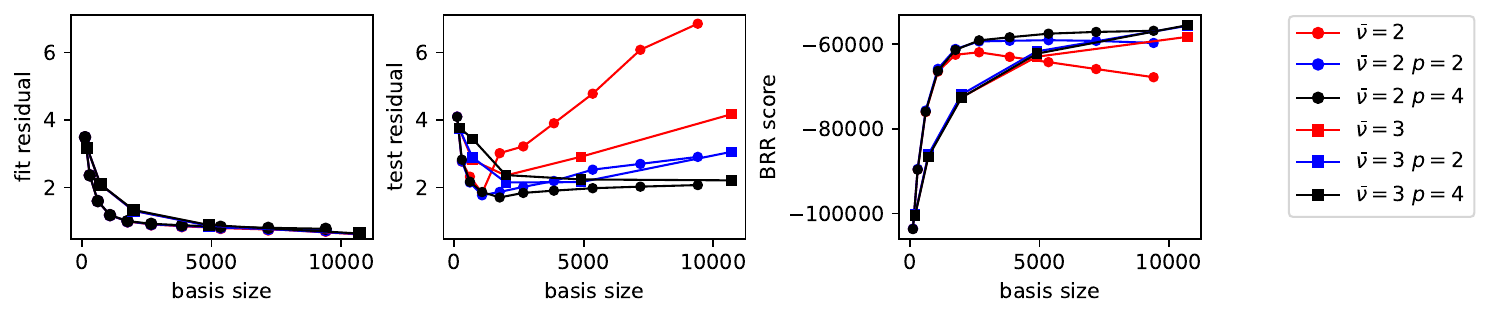}
    \caption{Fitting set residual (left), testing set residual (center), and log marginal likelihood (right) as a function
    of basis size for CsPbBr$_3$ ACE model fit to a database generated with HAL.
    Symbol indicates correlation order $\bar{\nu}$, and color indicates smoothness prior exponent $p$.}  
    \label{fig:perovskite_fit_test}
\end{figure*}

The training set residuals, test set residuals, and {\tt BayesianRidge} score (log marginal 
likelihood) are plotted as a function of basis size in Fig.~\ref{fig:perovskite_fit_test}. For each value of $\bar{\nu}$
the fitting error improves monotonically as the basis size (and polynomial degree) increases, but at equal basis 
size the $\bar{\nu}=2$ residuals are lower by as much as 25\% (especially for moderately sized
bases), indicating that for this system increasing the
polynomial degree provides the basis with more useful flexibility as compared to increasing  $\bar{\nu}$. 
For the basis size range where the error is minimized, the testing set residuals are larger than
the fitting set by at least about a factor of 2, indicating that some amount of overfitting is occurring.  The smoothness
prior is successful at limiting the extent of this overfitting. 

The generally lower training and test errors for the $\bar{\nu}=2$ models relative to the correlation
order three models are reflected in their Bayesian ridge scores (log marginal likelihoods). However, within each 
correlation order the optimal choice of polynomial degree and corresponding basis size indicated by 
the minimum test error are not consistent with the score. 
Indeed, the results displayed in Figure~\ref{fig:perovskite_fit_test} lead us to conclude that the Bayesian ridge score is not always a reliable tool for optimal basis selection and other options should be explored in the future. 

\begin{figure}
    \centering
    \includegraphics[width=0.5\columnwidth]{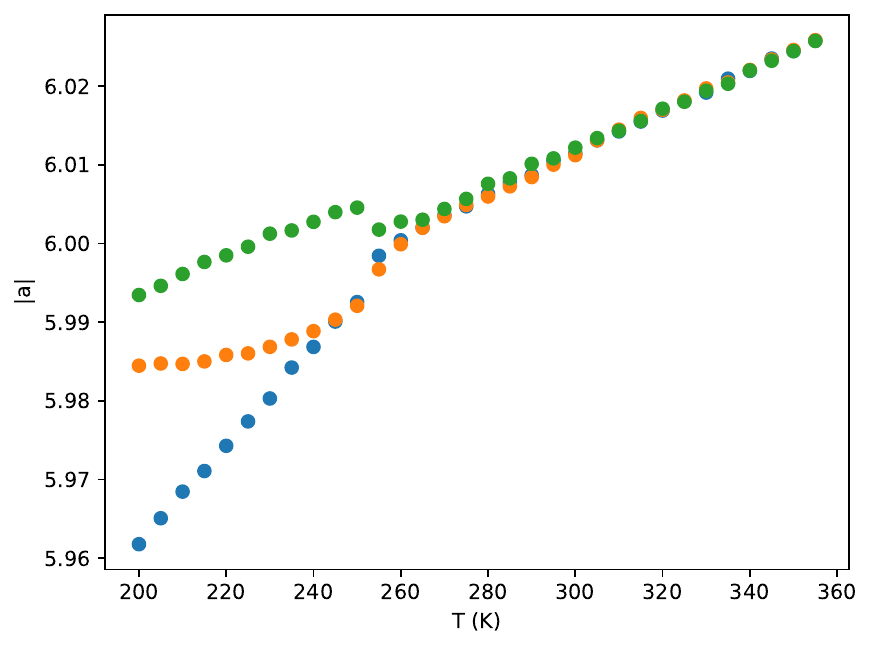}
    \caption{Effective cubic lattice constants at fixed temperature simulated using the ACE model
    with $\bar{\nu}=2$, maximum polynomial degree 12, and $p=4$. All
    three values are identical (to within the estimated error) at $T > 255$~K indicating a cubic structure.
    At lower temperatures these split into a single value and a group of two, consistent with a tetragonal
    structure, and at $T < 240$~K they split further into three distinct values, consistent with an orthorhombic
    structure.}  
    \label{fig:perovskite_phase_trans}
\end{figure}

We used the model with lowest test set error, generated by the fit with $\bar{\nu}=2$, maximum polynomial degree 12, and
smoothness prior $p=4$, to simulate larger unit cells of CsPbBr$_3$ at a range of temperatures
spanning its expected range of phase transition temperatures. We simulated 32 independent constant temperature, 
constant pressure, MD trajectories at temperatures from 200~K to 355~K and zero pressure for 10$^4$ 10~fs time steps.
Each trajectory started from an $8 \times 8 \times 6$ supercell (7680 atoms) of the orthorhombic structure.
To analyze the resulting structure we reconstructed the effective cubic lattice vectors and averaged their
magnitudes over the last 8000 steps of each trajectory. A plot of these effective cubic cell lattice vector
magnitudes as a function of temperature is shown in Fig.~\ref{fig:perovskite_phase_trans}. 
We see the three expected phases as indicated by the degeneracy of the lattice constants: cubic at high temperature, tetragonal at intermediate temperatures, and orthorhombic at low temperatures. 
The transition temperatures are 240~K and 255~K, which are substantially shifted relative to the experimental results of 361~K and 403~K~\cite{2013_stoumpos}.  
We expect that
this deviation from experiment is primarily due to our choice of exchange correlation functional, the Perdew-Burke-Enzerhof generalized-gradient approximation,~\cite{PBE} as has been seen in similar simulations~\cite{2023_fransson}. 
A direct comparison to DFT would be useful, but it would require an accurate calculation of the predicted phase transition temperatures directly from the DFT PES, 
which is too computationally demanding to be practical without additional approximations.

\section{Computational Performance}

\label{sec:computational_performance}

The linearity of ACE potentials renders them not only interpretable but also efficient in terms of computational performance. 
To demonstrate this, a performance test was conducted on various linear ACE potentials referenced in this paper. The evaluation times, as well as some ACE hyperparameters used, are shown in Table~\ref{tbl:performance_table} for the AlSi10, CsPbI$_{3}$, H$_2$O, PEG and Si potential developed in this work.
The number of basis functions for each model is given too and may be fewer than a complete ACE basis parameterized by $\bar{\nu}$ and $D^{\max}$ due to ARD pruning basis functions with low relevance. 
The timings were obtained using the LAMMPs-PACE implementation \cite{performant2022lysogorskiy} using a 128 core ARCHER2 node, equivalent to two seperate AMD EPYC 7742 64-core at 2.25GHz. 
The $10^6$ steps/day figures are equivalent to a ns/day and were obtained for varying cell sizes to illustrate scaling.
A standardized performance figure in the form of core-$\mu$s/atom figure is also provided.
\revise{The silicon database fitted originates from the silicon GAP potential, whereas the AlSi10, PEG and CsPbBr$_{3}$ potentials were fitted using HAL generated databases containing 22, 68 and 198 configurations respectively as discussed in the previous subsections.}

\begin{table}[ht]
\begin{tabular}{c|cccc|cc}

& \multicolumn{4}{c|}{ACE parameters}  & \multicolumn{2}{c}{Performance} \\
\hline
 &  $\bar{\nu}$ & $D^{\max}$ & $r_{\textrm{cut}}$ & \# basis func. & $10^6$ steps/day [atoms] & core-$\mu$s/atom \\
\hline
 AlSi10 & 2 & 7  & 5.5 & 79 & 636 [32] & 23 \\
 CsPbBr$_{3}$ & 2 & 12 & 5.5 & 544 & 334 [20] & 93 \\
 PEG & 3 & 12 & 5.5 & 4897 & 10 [1400] & 227 \\
  Si & 4 & \revise{20} & \revise{6} & \revise{5434} & \revise{7} [250] & \revise{744} \\
\hline
\end{tabular}
\caption{Performance of linear ACE potentials for various systems using an ARCHER2 node utilising 128 cores for the $10^6$ steps/day figures (equivalent ns/day using a 1 fs timestep). Core-$\mu$s/atom figures were obtained by performing simulations in serial.}
\label{tbl:performance_table}
\end{table}

\section{Conclusion and Outlook}
We introduced {\tt \revise{ACEpotentials}.jl}, a front-end for several Julia-language packages that implement Atomic Cluster Expansion (ACE) MLIPs and related functionality. This front-end provides a user-oriented interface, while the backend packages combine excellent performance with the flexibility for rapid model development and experimentation that is typical for the Julia language. The front-end {\tt \revise{ACEpotentials}.jl} exposes a relatively simple subset of ACE type models, linear models with robust priors, that we consider reliable in every-day use, especially in the context of an active learning type workflow.  

However, we emphasize that the ACE framework allows for a much richer MLIPs design space~\cite{2022-Bochkarev-mlace,performant2022lysogorskiy,MACE2022,ACE_ralf,darby2022compressing} as well as parameterisation of many other types of particle systems~\cite{2021-acetb1,2022-bips1,2022-wave1,2023-mlvmcace1}. We therefore conclude by mentioning some of those extensions, as well as current short-comings, that require further development. 

\begin{itemize} 
    \item Robust parameter estimation, in particular hyperparameter tuning, remains under-investigated in the MLIPs context. We regularly experience that hand-tuned hyperparameters can give superior results, basis sparsification remains poorly understood, and uncertainties are often only indicative of actual errors. Further research is required to resolve these closely related issues. 
    \item The design space of the {\tt \revise{ACEpotentials}.jl} ACE models can be expanded to admit trainable radial embeddings, composition of ACE features with nonlinearities, or even multi-layer architectures such as \cite{2022-Bochkarev-mlace,MACE2022}. This comes at the cost of highly nonlinear and less efficient models, but some of those extensions, such as trainable radial embeddings, can be undertaken while keeping the spirit of our current ACE models: small models for rapid iterative development and low evaluation cost.
    \item The extension to highly nonlinear models would likely require that the computational kernels on which {\tt \revise{ACEpotentials}.jl} is built also be made GPU-capable. Towards that end a deep learning framework such as MACE~\cite{MACE2022} (see also the {\tt mace}\footnote{\tt https://github.com/ACEsuit/mace} code) may be better suited. 
    \item Finally, we note that there are already several related ACE software packages within {\tt ACEsuit}\footnote{{\tt https://github.com/ACEsuit}} that implement a variety of models for other particle systems at different stages of development: Hamiltonians (\cite{2021-acetb1}, {\tt ACEhamiltonians.jl}); 
    wave functions (\cite{2022-wave1,2023-mlvmcace1}, {\tt ACEpsi.jl}); jet tagging models (\cite{2022-bips1}, {\tt BIPs.jl}). These build on an experimental and significantly expanded Julia-language ACE package {\tt ACE.jl}. 
\end{itemize}

\begin{acknowledgements}
    GC acknowledges support from EPSRC grant
EP/X035956/1. CO, AR and TJ were supported by NSERC Discovery Grant GR019381 and NFRF Exploration Grant GR022937. WB was supported by US AFRL grant FA8655-21-1-7010. C vd O and GC acknowledge ARCHER2 for which access was obtained via the UKCP consortium and funded by EPSRC grant EP/P022065/1. NB was supported by the U.~S. Office of Naval
    Research through the U.~S. Naval Research Laboratory's fundamental research base program. EG acknowledges support from the EPSRC Centre for Doctoral Training in Automated Chemical Synthesis Enabled by Digital Molecular Technologies with grant reference EP/S024220/1.
    WCW was supported by the Schmidt Science Fellows in partnership with the Rhodes Trust, and additionally acknowledges support from EPSRC (Grant EP/V062654/1).
    JRK and CO acknowledge funding from the Leverhulme Trust under grant RPG-2017-191 and the EPSRC under grant EP/R043612/1. JRK\revise{, JPD} and GC acknowledge support from the NOMAD Centre of Excellence funded by the European Commission under grant agreement 951786. JRK acknowledges support from the EPSRC under grants EP/P002188 and EP/R012474/1.
    This work was performed using resources provided by the Cambridge Service for Data Driven Discovery (CSD3) operated by the University of Cambridge Research Computing Service (www.csd3.cam.ac.uk), provided by Dell EMC and Intel using Tier-2 funding from the EPSRC (capital grant EP/T022159/1), and DiRAC funding from the STFC (www.dirac.ac.uk). \revise{Further computing facilities were provided by the Scientific Computing Research Technology Platform of the University of Warwick.}
    
    \textit{For the purpose of open access, the corresponding author has applied a Creative Commons Attribution (CC BY) licence to any Author Accepted Manuscript version arising from this submission.}
\end{acknowledgements}

\appendix

\section{Linear Scaling Cost and Computational Kernels}
\label{sec:implementation}
In Sections \ref{sec:model:ace_framework} and \ref{sec:model:basis} we outlined some basic ideas behind the ACE model, in particular expressing the potential energy model in terms of the many-body expansion \eqref{eq:manybody}. A naive implementation of the many-body expansion results in prohibitive computational cost due to the exponential cost of the sums over clusters $(j_1, \dots, j_\nu)$. However, after discretizing the $U^{(\nu)}$-body potential of the {\em self-interacting many-body expansion} \eqref{eq:manybody_dirty} the sum can be rewritten to result in linear scaling cost. This is presented in detail, for example, in~\cite{ACE_ralf, performant2022lysogorskiy, DUSSON2022}, hence we shall not review this process in full detail here. In order to outline what is involved in an implementation of an ACE potential, we only recall the form that \revise{the} ACE model takes after this re-organisation of the many-body summation. The evaluation of the {\em self-interacting} ACE basis then results in the following stages:

\begin{enumerate}
    \item Evaluation of the embeddings, $R_{nl}(r_{ij}, Z_i, Z_j)$ and $Y_l^m(\hat\rr_{ij})$.
    \item A pooling operation; also called called the atomic basis~\cite{ACE_ralf}, or the density projection~\cite{bartok2010gaussian}, 
    \begin{equation}
        A_{znlm}^i = \sum_{j \in \calN{(i)}} \phi_{znml}(\rr_{ij}, Z_j, Z_i),
    \end{equation}
    \revise{where $\mathcal{N}(i)$ denotes the set of indices of all atoms within the cutoff radius from atom $i$.}
    \item Product basis: for lexicographically ordered tuples $(\zz,\nn,\ll,\mm) = (z_t, n_t, l_t, m_t)_{t = 1}^\nu$ we define
    \begin{equation}
        \bAA_{\zz \nn \ll \mm}^i = \prod_{t = 1}^\nu A^i_{z_t n_t l_t m_t}.
    \end{equation}
    This operation can be thought of as a sparse symmetric tensor product, or as taking $\nu$-correlations.
    \item Symmetrization: To ensure invariance one averages $\bAA^i$ over all rotations, resulting in the $O(3)$-invariant basis 
    \begin{equation} \label{eq:BB_dirty}
        \BB^i = \mathcal{C} \bAA^i, 
    \end{equation}
    employed in the definition of the linear ACE model~\eqref{eq:lin_ace}. Here, $\bAA^i$ is the vector of $(\bAA_{\zz \nn \ll \mm}^i)$ basis functions while $\mathcal{C}$ a sparse matrix.     
\end{enumerate}

For each of these stages efficient computational kernels are implemented, designed in a modular way so that they can be independently optimized or composed into new model architectures.

\subsubsection*{Canonical Many-Body Expansion}
Under the condition that the radial basis and envelope function are pure polynomials, it is possible to transform the self-interacting ACE basis $\BB^i$ defined in \eqref{eq:BB_dirty} into a basis for the canonical many-body expansion \eqref{eq:manybody_pure}. The idea behind this procedure is sketched out in \cite{DUSSON2022}. The precise details of the implementation and a detailed study is not the purpose of this review. 
Here, we only mention that, upon slightly extending the $R_{nl}, A^i$ and $\bAA^i$ bases, one can obtain a ``purification operator'' $\mathcal{P}$ such that the linearly transformed $\mathcal{P} \bAA^i$ becomes a basis for the canonical many-body expansion \eqref{eq:manybody_pure}. 
The symmetrisation $\mathcal{C}$ can then be applied to obtain an $O(3)$-invariant basis $\mathcal{B}^i := \mathcal{C} \mathcal{P} \bAA^i$. 

An important variation of the ``purification operation'' $\mathcal{P}$ is to only purify the 2-body interaction. This entails replacing the fully self-interacting basis functions
\[
    \bAA^i_{\kk} = \sum_{j_1, \dots, j_\nu} 
            \prod_{t=1}^\nu \phi_{k_t}(x_{i j_t})
    \qquad \text{with} \qquad 
    \sum_{\substack{j_1, \dots, j_\nu \\ j_a \neq j_b}} 
        \prod_{t=1}^\nu \phi_{k_t}(x_{i j_t})
\]
All three options (i) fully self-interacting, (ii) purified pair interaction, and (iii) canonical cluster expansion are available in {\tt \revise{ACEpotentials}.jl}. The package documentation should be reviewed on how to select the different basis sets.

\bibliographystyle{unsrt}
\bibliography{references}

\end{document}